\documentclass[10pt,twocolumn]{article}

\usepackage[sc]{mathpazo} 
\usepackage[T1]{fontenc}
\usepackage{float}
\usepackage{subfig}
\usepackage{textcomp}
\usepackage[hmarginratio=1:1,left=20mm,right=20mm,top=20mm,columnsep=20pt]{geometry}

\usepackage{graphicx}
\usepackage{mathptmx}      
\usepackage{flushend}
\usepackage{hyperref}
\usepackage[numbers,sort&compress]{natbib}

\usepackage{abstract}

\newcommand{\abs}[1]{\left| #1 \right|} 
\newcommand{\avg}[1]{\left< #1 \right>} 

\title{\vspace{5mm}%
	\fontsize{20pt}{10pt}\selectfont
	\textbf{3D Particle Track Reconstrution in a Single Layer Cadmium-Telluride Hybrid Active Pixel Detector}
	}	
\author{%
	\large Mykhaylo Filipenko\thanks{mykhaylo.filipenko@physik.uni-erlangen.de}, Thomas Gleixner, Gisela Anton and Thilo Michel  \\[2mm]
	\normalsize	Erlangen Center for Astroparticle Physics, University of Erlangen \\
	\normalsize Erwin-Rommel-Stra\ss e 1 \\
	\normalsize 91058 Erlangen, Germany
}
\date{\today}

\usepackage{titlesec}
\renewcommand\thesection{\Roman{section}}
\titleformat{\section}[block]{\large\scshape\centering}{\thesection.}{1em}{}

\begin{document}

\maketitle

\begin{abstract}
In the past 20 years the search for neutrinoless double beta decay has driven many developements in all kind of detector technology. A new branch in this field are highly-pixelated semiconductor detectors - such as the CdTe-Timepix detectors. It compromises a cadmium-telluride sensor of 14 mm x 14 mm x 1 mm size with an ASIC which has 256 x 256 pixel of 55 \textmu m pixel pitch and can be used to obtain either spectroscopic or timing information in every pixel. In regular operation it can provide a 2D projection of particle trajectories; however, three dimensional trajectories are desirable for neutrinoless double beta decay and other applications. In this paper we present a method to obtain such trajectories. The method was developed and tested with simulations that assume some minor modifications to the Timepix ASIC. Also, we were able to test the method experimentally and in the best case achieved a position resolution of about 90 \textmu m with electrons of 4.4 GeV.
\end{abstract}





\section{Introduction}

The main motivation behind the methods and experiments presented in this paper is the demonstration of the possiblity of 3D particle trajectory reconstruction within the sensor of a hybrid active pixel detector for a neutrinoless double beta experiment with high sensitivity due to background rejection by particle tracking. The neutrinoless double beta decay is a hypothetical, lepton-number violating decay where two neutrons in a nucleus are transformed into two protons and two electrons without the anti-neutrino emission, e.g.

\begin{eqnarray}
	^{116}\textrm{Cd} \rightarrow ^{116}\textrm{Sn} + 2e^{-}
\end{eqnarray}

This decay is predicted within the framework of Majorana-neutrinos, introduced by Ettore Majorana in 1937 \cite{majorana}, which is a possible extension to the standard model of particle physics with extensive consequences: Within this framework it can be deduced that the neutrino is  its own anti-particle and has a mass different from zero (in contrast to the standard model where the neutrino is massless). A non-zero neutrino mass is strongly enforced by evidence from neutrino oscillation experiments. Furthermore, up to now only bosons are known to be their own anti-particles and the neutrino would be the only fermion to fulfill this property. Therefore, besides the direct evidence for lepton-number violation, the observation of neutinoless double beta decay would have an immense impact on our understanding of particle physics \cite{dbreview}.\par
Although the search for the neutrinoless double beta decay is the main reason for our investigations, semiconductor voxel detectors could have other interesting applications as well. When using a thin sensor coupled to a pixelated ASIC,the read-out volume of single pixel is a tick with pixel area as ground area and the sensor thickness as height, in our case 55 x 55 x 1000 \textmu m$^3$. In order to resolve 3D particle tracks inside the sensor volume it is necessary to obtain position resolution along the height of the sensor. We name a detector which is able to perform this task, a voxel detector. \par
As high resolution voxel detectors, semiconductor detectors could be used for efficient high energy single photon Compton-imaging. By reconstructing the electron track not only the energy of an X-ray photon can be determined but also the direction of its origin. Further, it would allow the usage for low activity tracers in SPECT imaging as the the collimator, which absorbs most of the flux, could be avoided. Additionally, Compton-imaging could be used for security application; for instance, if a method for fast and precise detection of nuclear contamination is needed \cite{kailbnl}, \cite{msu}. \par
3D imaging of particle tracks could also allow to perform a variaty of other important low-background experiments like the observation of the double electron capture. Another important experiment that one could aim at with voxel detectors is the precise measurement of the angle between the direction of emittion of high energy photons from the consequent deexcitation of energy states of the nucleus into ground state \cite{physrevbenni}. Summing up, the development of voxel semiconductor detectors would enhance the possiblities of detection in various fields.\par
In this paper we present a method for the reconstruction of such 3D trajectories through the sensor layer. It was deleveloped and optimized with simulations. Also, the method could successfully be applied to experimental data obtained by a Timepix detector \cite{mpx} and used for the reconstruction of electron tracks through the sensor layer. \par
The paper is structured as following: In the second section the Timepix detector is explained as our simulations and experiments are based on this technology. In the third section the reconstruction method is described and in the fourth section the experimental results are presented. Section V gives an outlook on the current detector development which could lead to a fully operationable voxel detector with about 50 \textmu m voxel pitch.

\section{The Timepix Detector}

The Timepix detector is a pixelated hybrid semiconductor detector which was originally developed for X-ray imaging applications \cite{mpx}. Hybrid in this context means that the sensitive sensor layer and the ASIC are fabricated seperately and afterwards connected together via bump-bonds. This has the advantage that different semiconductor materials can be used for the sensor layer. For X-ray imaging it is desirable to have sensor material with a high mass number in order to achieve high absorption efficiency. Hence, besides silicon cadmium-telluride was established as a sensor material for Timepix devices. Despite the difficulty of obtaining large high quality cadmium-telluride crystals the advantage of room-temperature operation and a mass number 48 and 52, respectively, makes this material preferable (to Germanium) for X-ray imaging applications.\par

\begin{figure}[tb]
\centering
\includegraphics[width=\columnwidth]{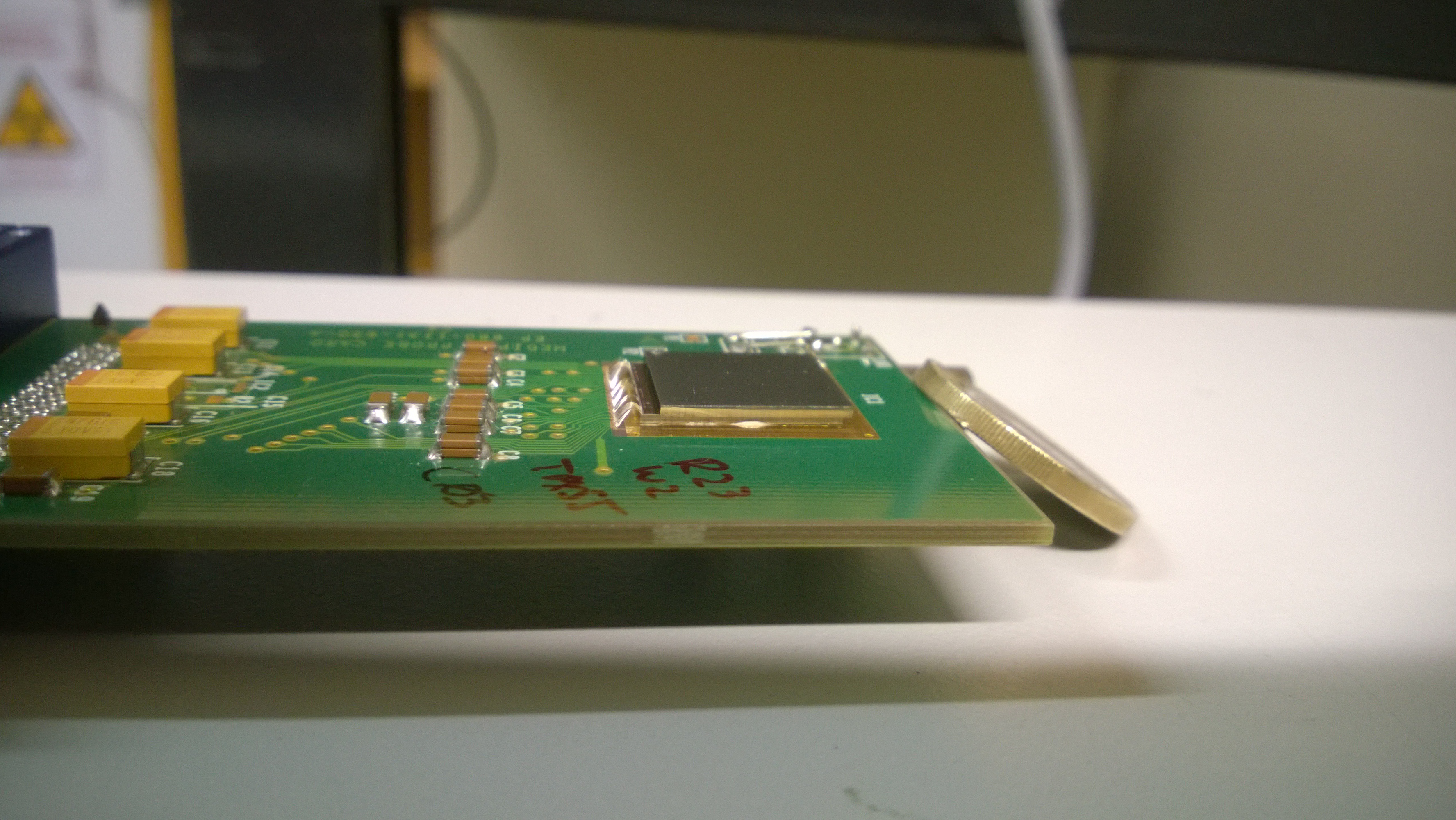}
\caption{A Timepix detector with a 1mm thick cadmium-telluride sensor.}
\label{timepix_pic}
\end{figure}

\begin{figure}[tb]
\centering
\includegraphics[width=\columnwidth]{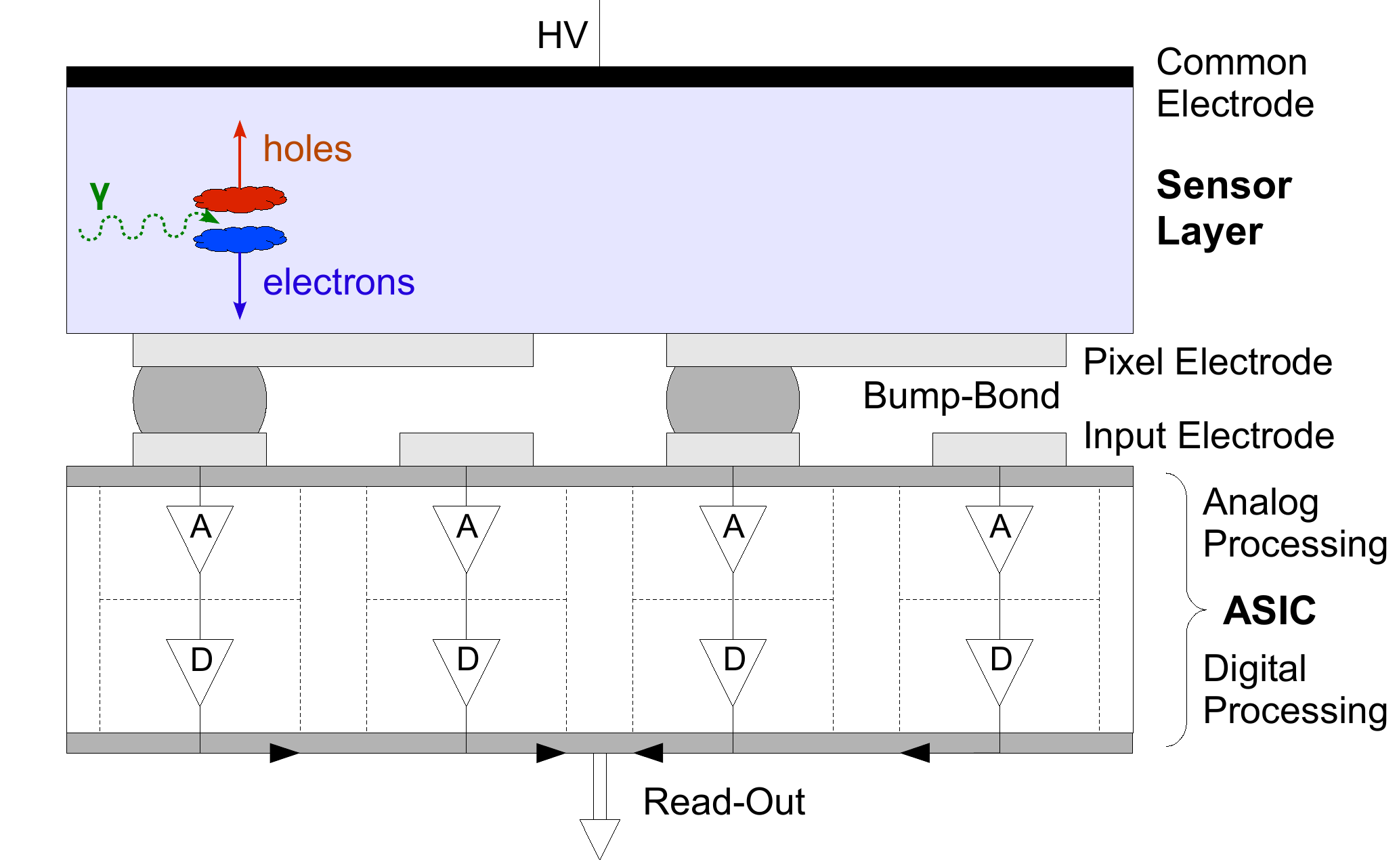}
\caption{A cut through a Timepix-detector with a CdTe sensor layer. A high-voltage is applied to the sensor volume. The upside is at HV, the downside at ground potential. Electrons are excited by ionizing radiation from the valence into the conduction band and drifted by the electric field. During the drift the electrons induce charge at the pixel electrodes which is amplified and converted to a voltage pulse by the electronics in the ASIC.}
\label{timepix_scheme}
\end{figure}

A scheme of the Timepix structure and functionality is shown in Fig.\ref{timepix_scheme}. The Timepix used in this study consists of a 1 mm thick cadmiun-telluride sensor with platinum electrode pads (ohmic contacts) as pixel electrodes at the bottom and a common electrode for the high-voltage connection at the top. The sensitive sensor layer is connected pixel-by-pixel to the input pixel electrodes of the ASIC by indium bump-bonds. The pitch of the input pixel electrodes is 55 \textmu m. There are 256 x 256 pixels on the ASIC which corresponds to an active area of 1.4 x 1.4 $\textrm{ cm}^2$. In our detector only every second pixel in each row and column is bump-bonded wherefore the effective pixel size is 110 \textmu m. This has the advantage of better energy resolution at the cost of position resolution. The ASIC is mounted on a printed circuit board and connected to the computer by a USB-FitPix readout \cite{fitpix} for the data transfer to the data acquisition computer which can provide up to 60 frames per second (s. Fig. \ref{timepix_pic}). The output data is provided in frames. Each frame is a 128 x 128 matrix of values obtained for every pixel during the frame-time. The frame-time can be preset manually to a fixed time or controlled externally by a trigger. \par

The sensor layer can be biased with a voltage up to 800 V. Secondary electrons and holes generated by ionizing radiation are separated and afterwards drifted by an eletrical drift field in opposite directions. While approaching the pixel electrode the charge cloud induces mirror charges in the pixel electrodes. In each pixel the induced charge is collected and converted into a triagular shapped pulse by a Krummernacher type pre-amplifier with an approximate rise time of about 100 ns. The length of the pulse is about several microseconds which is mainly determined by the falling edge of the pulse. The gain and the falling edge of the pre-amplifier can be set externally by a digital-to-analog converter in the matrix periphery. The pulse in each pixel is discriminated against a global threshold.\par
\begin{figure}[tb]
\includegraphics[width=0.95\columnwidth]{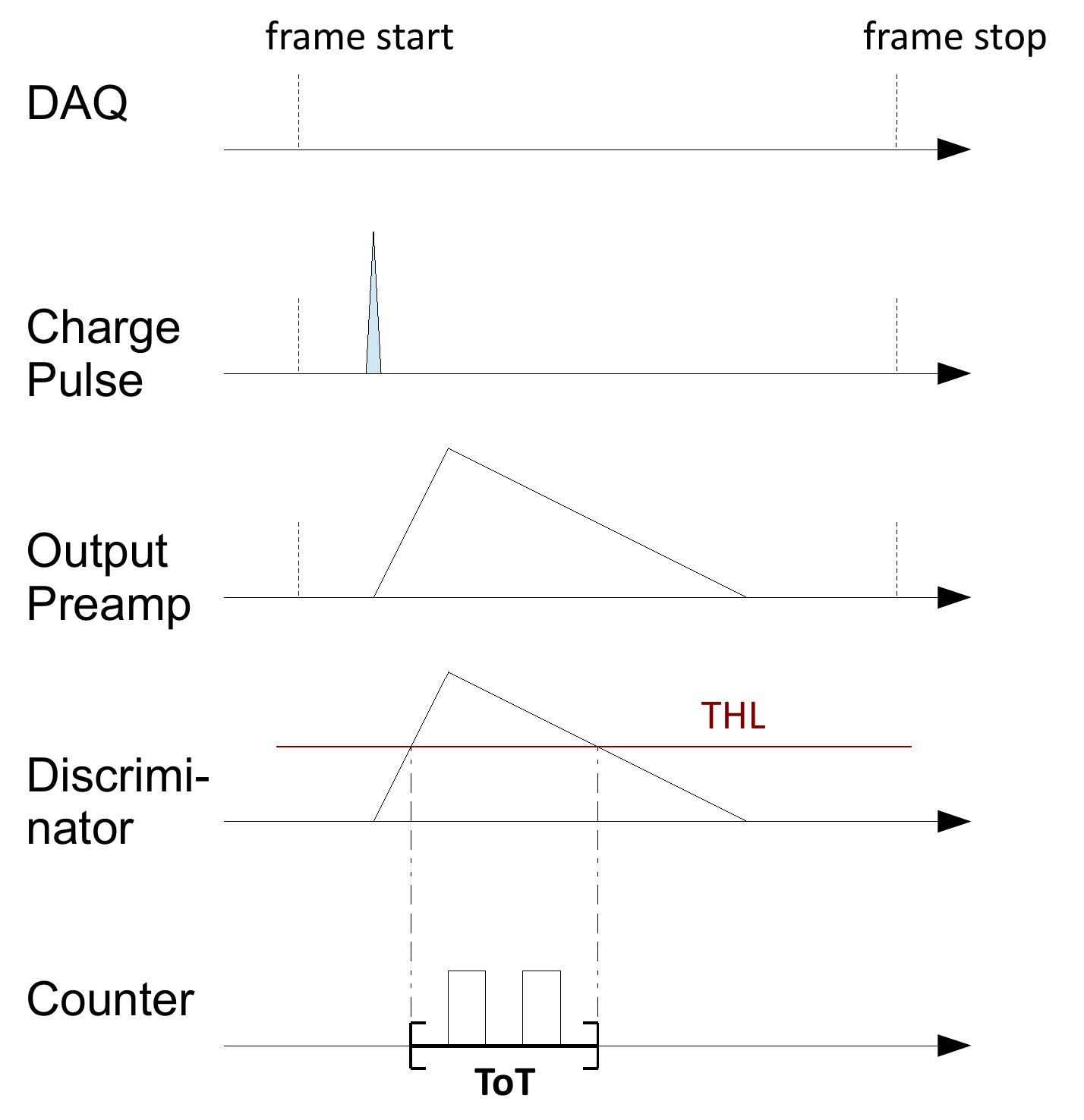}
\caption{The operation of the Timepix in the "time-over-threshold"' (ToT) mode. After the charge pulse is amplified and converted to a voltage pulse, it is discriminated against a threshold level (THL). The time-over-threshold is the number of clock counts in the time interval of the pulse above the THL.}
\label{timepix_tot}
\end{figure}
\begin{figure}[tb]
\includegraphics[width=0.95\columnwidth]{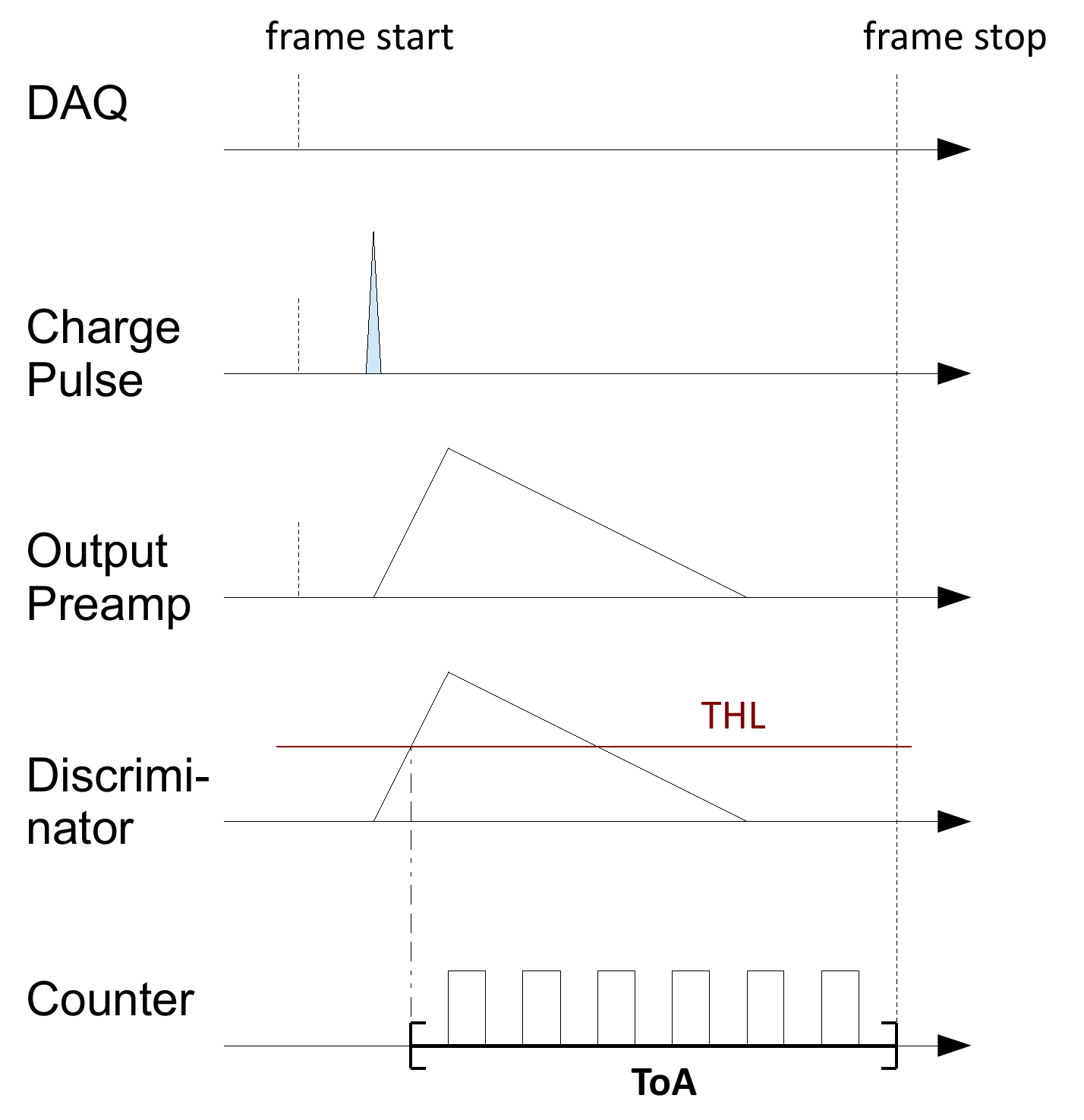}
\caption{The operation of the Timepix in the "time-of-arrival"' (ToA) mode. After the charge pulse is amplified and converted to a voltage pulse, it is discriminated against a threshold level (THL). The time-of-arrival is the number of clock counts between the first time the voltage pulse rises over the THL and the end of the frame.}
\label{timepix_toa}
\end{figure}

The electric post-processing of the signal in every pixel is digital and can be carried out in three different ways. We used the detector only in two modes - the "`time-over-threshold"' (ToT) mode and the "`time-of-arrival"' (ToA) mode. Both modes are illustrated in Fig. \ref{timepix_tot} and Fig. \ref{timepix_toa}, respectively.\par

In the ToT mode a digital register starts counting clock cycles when the pulse rises over the threshold. The register stops once the pulse falls below the threshold and the number of counts inbetween is the time-over-threshold. The frequency of clock cycles can be chosen between 10 MHz and 100 MHz. The time-over-threshold is related to the height of the pulse which is a measure for the energy deposited in a pixel. Thus, the time-over-threshold can provide a calorimetic information about a detected event. The deposited energy and the ToT are related in a non-linear way. The energy resolution achievable in this mode is $8.0 \%$ (FWHM) at 59.56 keV and $5.4 \%$ (FHWM) at 122.06 keV. Details on the calibration and the energy resolution can be found in \cite{calibpub}. \par
In the ToA mode the register starts counting on the rising edge of the pulse and stops only at the end of the frame. As every frame gets a global time stamp, the ToA provides a timing information for every event. It can be as good as 10 ns for every pixel if a 100 MHz clock is used. The maximum counter value is 11280 which corresponds to a frame-time of $112.8 \textrm{ \textmu s}$ at a 100 MHz clock frequency.

\section{Reconstruction Method}

\begin{figure}[tb]
\subfloat[$t = t_0$]{\includegraphics[width=\columnwidth]{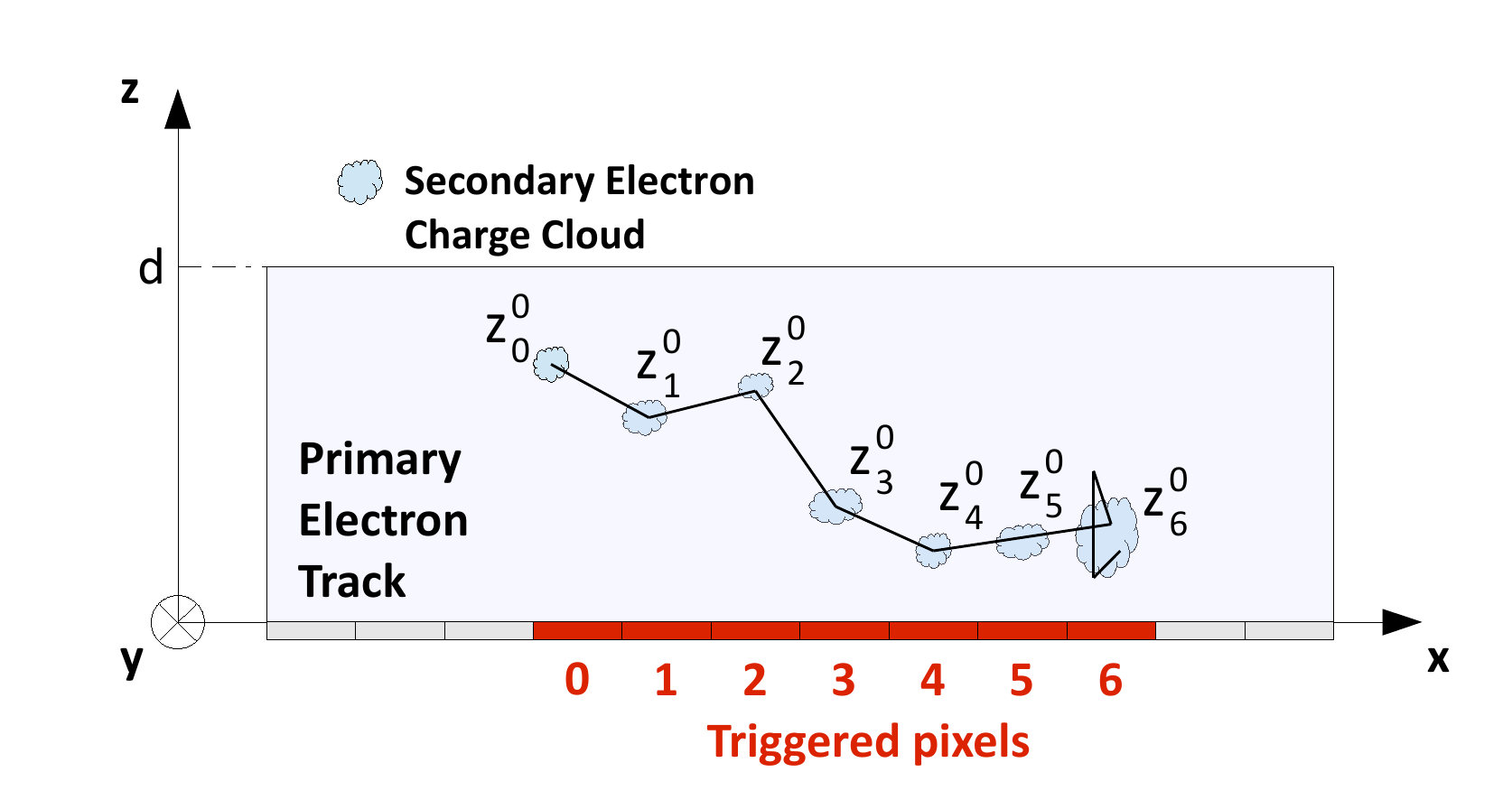}}

\subfloat[$t = t^{THL}_6$]{\includegraphics[width=\columnwidth]{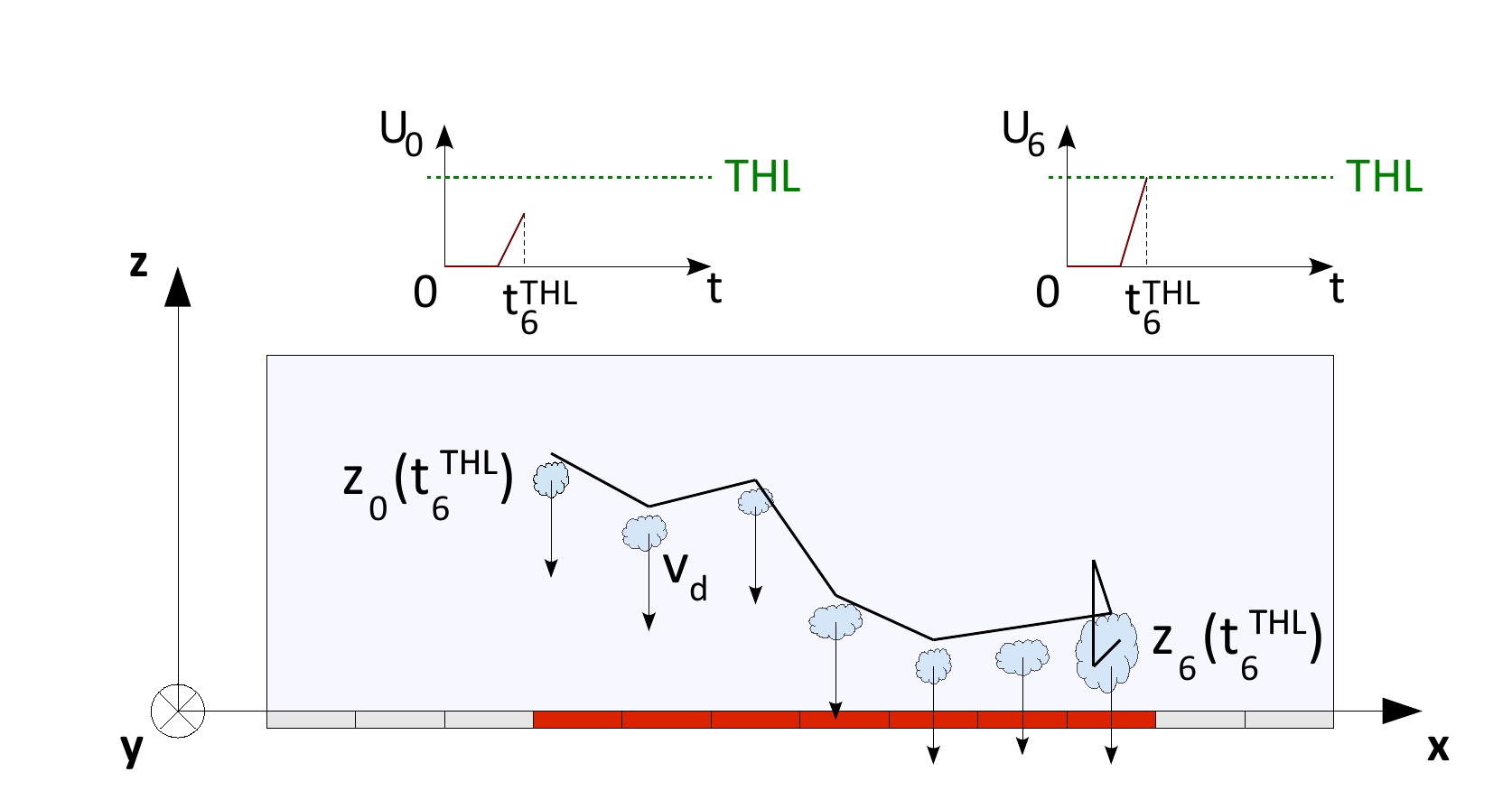}}
\caption{A cut through a pixelated detector which illustrats the goal of the reconstruction. The depth of interaction $z^0_n$ should be reconstructed for every pixel. The triggered pixels are red. a) The event happens at $t_0$; b) a snapshot at $t = t^{THL}_6$: All charge clouds were drifted towards the pixel electrodes. In pixel 6 the charge cloud is close enough to the electrode for the voltage pulse ($U_6$) to go over the threshold (THL). At the same time the voltage pulse in pixel 0 ($U_0$) has not reached the threshold yet.}
\label{reconstruction_scheme}
\end{figure}

In order to reconstruct the three-dimensional structure of particle trajectories through the sensor material, it is neccesary to determine the depth of interaction $z^0_n$ in the sensor for every triggered pixel $n$ (Fig. \ref{reconstruction_scheme}). Without the loss of generality, we assume that the sensor is pixelated in the xy-plane and the depth of interaction is to be calculated along the z-axis. The sensor anode is at z = 0 and the cathode is at z = d; d being the sensor thickness. 

\subsection{Principle of Reconstruction}

The reconstruction is based on the charge carrier propagation mechanism in the sensor layer. Charge carriers are drifted in z-direction due to an electric field that can be approximated in a CdTe sensor layer by

\begin{eqnarray}
	\abs{E_z(z,U)} = U(f_2 + f_1 z + f_3 \exp(-f_4 U z)) \label{for_efield}
\end{eqnarray}

which is an emperical fit to experimental data \cite{hindawi}. The parameters $f_1$ to $f_4$ are $f_1 = (5.80 \pm 0.09) \cdot 10^{5} \ m^{-2}$, $f_2 = 228 \pm 7.5 \ m^{-1}$, $f_3 = 540 \pm 144 \ m^{-1}$ and $f_4 = 479 \pm 216\ (Vm)^{-1}$. U is the voltage difference applied to the sensor volume. The drift velocity for electrons can be calculated as $ v_z(z) = \mu_z E(z)$ where $\mu_z$ is the mobility of electrons in the z-direction ($\mu_z \approx 1100 \frac{cm^2}{Vs}$ in CdTe). Further, the interdependence between time and the z position of a charge carrier in the sensor is described by the following equation:

\begin{eqnarray}
	z(t) &=& z_0 + \int^{t}_0 E(z(t')) \mu_z \textrm{d} t' \\ \label{for_zpath}
			  &=& z_0 + \mu_z U \int^{t}_0 f_2 + f_1 z(t') + f_3 \exp(-f_4 U z(t')) \textrm{d} t'. \nonumber
\end{eqnarray}

Since this integral equation has no closed analytical solution for z(t), we used a numerical method to solve for z(t). We devided the total path between the anode and cathode into $20 000$ segments on which Eq. \ref{for_efield} can be approximated by a linear dependence on every segment i: $\abs{E(U,z)} = U(a_i + b_i\cdot z)$ (For every segment i, the parameters $a_i$ and $b_i$ are calculated). With this approximation Eq. 3 can be solved and we have a sufficient solution for z(t) on every segment.\par

As stated in the previous section the signal that is actually measured is the amount of charge that is induced in the pixel electrode by the charge cloud of secondary electrons during their drift through the sensor. We consider only the charge induced by the electrons and neglect the charge that is induced by the holes. The mobility of holes is $\mu_h \approx 100 \frac{cm^2}{Vs}$, hence more than 10 times lower than the mobility of the electrons. Also the weighting potential at the cathode is less steep than at the anode. Therefore, the main contribution of the holes to the signal happens after the integration time of the chip and can be neglected.\par

Suppose the charge $Q^{dep}_n$ was released at the time $t_0$ above the pixel electrode $n$ and is drifted from its initial position $z^0_n$ over a time period $t_d$. After this time periode, the amount of charge that is induced in the pixel electrode $Q^{ind}_n$ can be calculated as \cite{qindcalc}

\begin{eqnarray}
	Q^{ind}_n(t_d) = [W_{pot}(z_n(t_d)) - W_{pot}(z^0_n)]\ Q^{dep}_n. \label{for_qinduced}
\end{eqnarray}

\begin{figure}[tb]
\includegraphics[width=\columnwidth]{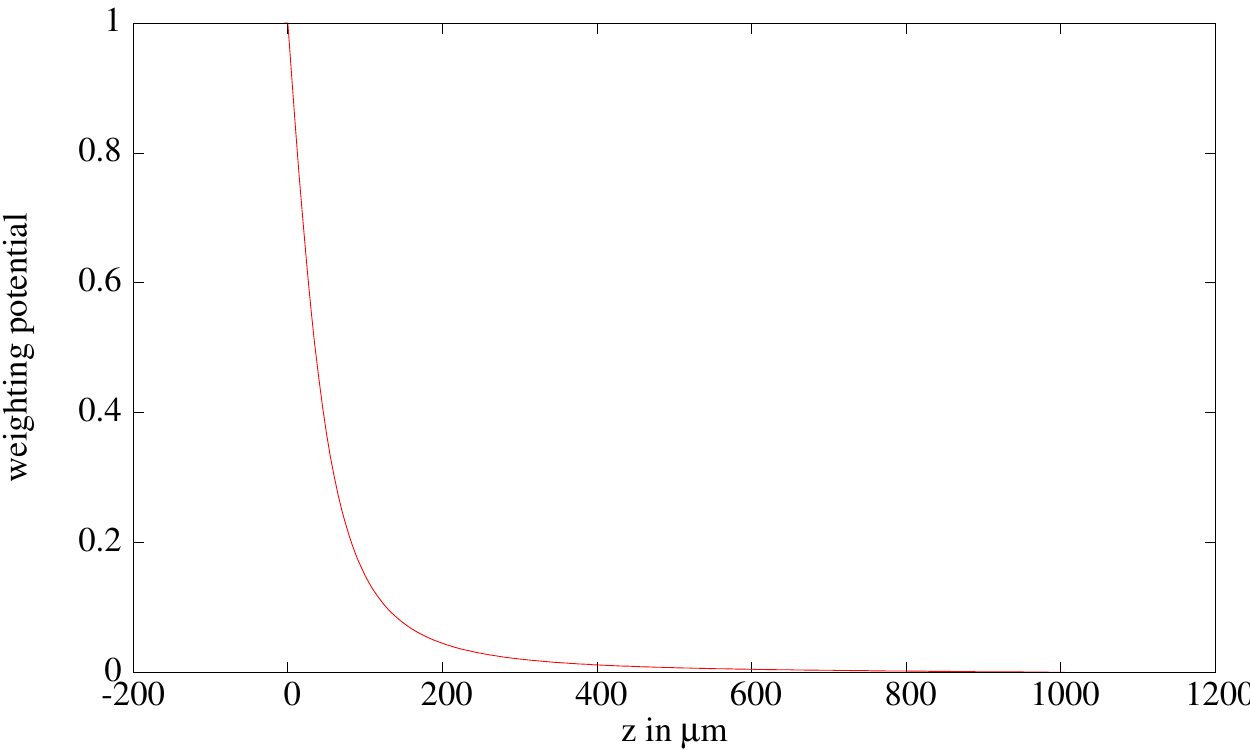}
\caption{The weighting potential in a CdTe-sensor with 110 \textmu m pixel size in z-direction calculated with the method of Castoldi \cite{weightpot}.}
\label{weight_pot}
\end{figure}

The function $W_{pot}$ is called the weighting potential and takes into account the geometric properties of the sensor with its pixel electrodes. It can be determined by solving the Laplace-equation for a given geometry \cite{weightpot}. For a geometry of rectangular shaped electronic pads the weighting potential along the z-axis is shown in Fig. \ref{weight_pot}. It was calculated with the method of Castoldi \cite{weightpot}.\par

For every pixel the threshold equivalent charge (or energy) is known after it has been determined in a calibration process. We will call this quantity $Q^{THL}_n$. Suppose this amount of charge was induced in the pixel after the time $t^{THL}_n$, then Eq. \ref{for_qinduced} yields 

\begin{eqnarray}
	Q^{THL}_n &=& Q^{ind}_n (t^{THL}_n)  \\ \nonumber
						&=&[W_{pot}(z(t^{THL}_n)) - W_{pot}(z^0_n)] \ Q^{dep}_n. \label{for_qind_thl}
\end{eqnarray}

If a detector could measure $t^{THL}_n$, the position $z^n_0$ could be determined by inverting Eq. 5. Since $z_n(t)$ depends also on $z^0_n$ (Eq. 3) in a non-trivial way, the equation cannot be solved analytically. Instead it can be solved numerically by iteration: We set $z^n_0$ to $\frac{d}{2}$ in the first step ($j = 1$) and for every pixel calculate the expression 

\begin{eqnarray}
\Delta Q_n(z^0_n) = Q^{THL}_n - [W_{pot}(z_n(t^{THL}_n)) - W_{pot}(z^0_n)] \ Q^{dep}_n. \label{for_z0condition}
\end{eqnarray}

In the case $\Delta Q > 0$ we increase $z^0_n$ by $\Delta z_{shift} = \frac{d}{2^j}$ where $j$ is the step number. In the other case ($\Delta Q < 0$) we decrease $z^0_n$ by $\Delta z_{shift}$. The deviation of the calculated position from the actual position is smaller then $\frac{d}{2^j}$ after the j-th step. Therefore, we used 30 steps in our analysis which produces a sufficient numerical results for $z^0_n$.\par 
However, an absolute measurement of $t^{THL}$ would require to measure the time of interaction of the incidenting particle with the sensor. With a semiconductor this is impossible since the charge signal is always delayed by the time that it takes to induce a signal at the pixel electrode; and this time - in turn - depends on the depth of interaction.\par

\begin{figure}[tb]
\includegraphics[width=\columnwidth]{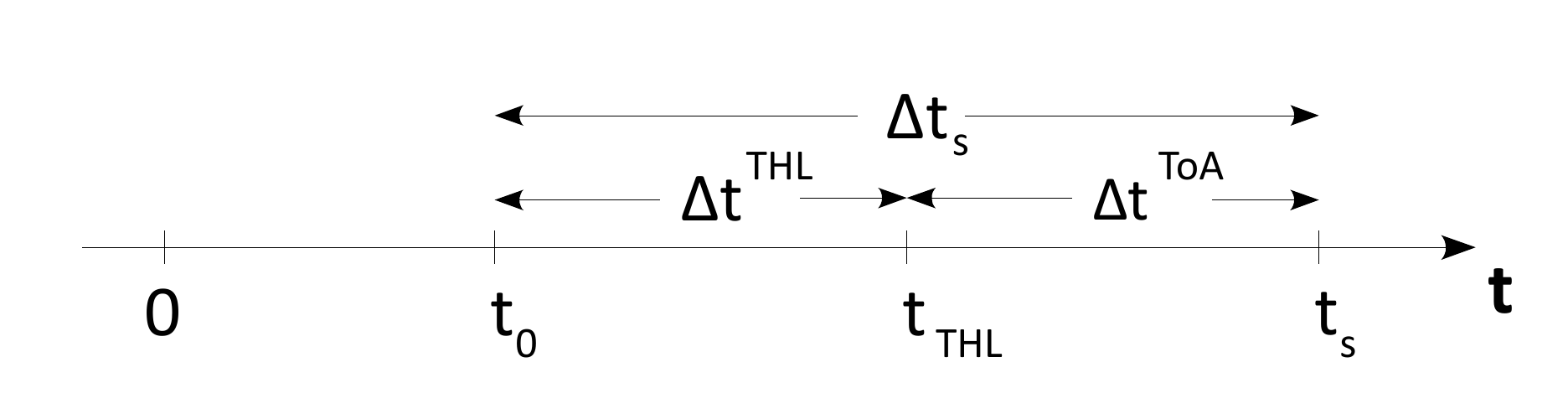}
\caption{An illustration of the time point and time intervals occuring during the drift process. The interaction of the ionizing particle with the sensor happens at $t_0$. At $t_{THL}$ the charge pulse (in a particular pixel) rises above the the threshold. At $t_s$ the frame stops.}
\label{reconstruction_timeline}
\end{figure}

What a detector can measure is a time-of-arrival for the charge signal in every pixel in the track relative to a fixed time-stamp set by the electronics $t_s$. We will call this quantity $\Delta t^{ToA}_n$; the time between $t_0$ and $t^{THL}_n$ is called $\Delta t^{THL}_n$; the time between $t_0$ and $t_s$ is referred to as $\Delta t_s$. This is illustrated in Fig. \ref{reconstruction_timeline}. According to Fig. \ref{reconstruction_timeline} $\Delta t^{THL}_n$ can be calculated as 

\begin{eqnarray}
	\Delta t^{THL}_n &=& \Delta t_s - \Delta t^{ToA}_n.
\end{eqnarray}

In the last formula $\Delta t_s$ is unknown but this quantity is the same for every pixel. By varying $\Delta t_s$ the center of gravity in z-direction (the z-value averaged over all pixels) is shifted since the $t^{ToA}$ values are fixed. It can be calculated as 

\begin{eqnarray}
	\avg{z} &=& \frac{1}{Q} \sum^{N}_{n = 0} z^0_n Q^{dep}_n
\end{eqnarray}

where $Q = \sum^{N}_{n=0} Q^{dep}_n$ is the total released charge in the interaction. Since the anode is not pixelated, the weighting potential at the anode $W_{pot}(0)$ is different from the weighting potential at the cathode $W_{pot}(d)$. As a consequence the ratio of the anode and cathode signal depends on the average depth of interaction \cite{msuzpos}. It can be determined after measuring the total induced charge at the anode $Q_a$ ($Q_c = Q$) and at the cathode $Q_c$ and comparing the $Q_c$/$Q_a$ ratio to a simulation-generated lookup-table.\par
If $\Delta t_s$ is assumed too high, then we get $\avg{z}^{meas} < \avg{z}$ because we overestimated the drift distance. In the opposite case we get $\avg{z}^{meas}>\avg{z}$ because we underestimated the drift distance. Therefore, $\Delta t_s$ can be determined in the same iterative precedure as $z^0_n$ under the condition

\begin{eqnarray}
	\Delta z &=& \avg{z}^{meas} - \avg{z} \\ \label{for_zavgcondition}
					 &=& \avg{z}^{meas} - \frac{1}{Q} \sum^{N}_{0} z_n(\Delta t_s) Q^{dep}_n. \nonumber
\end{eqnarray}

\subsection{Reconstruction Procedure}

\begin{figure}[tb]
\subfloat[Simulated Track]{\includegraphics[width=\columnwidth]{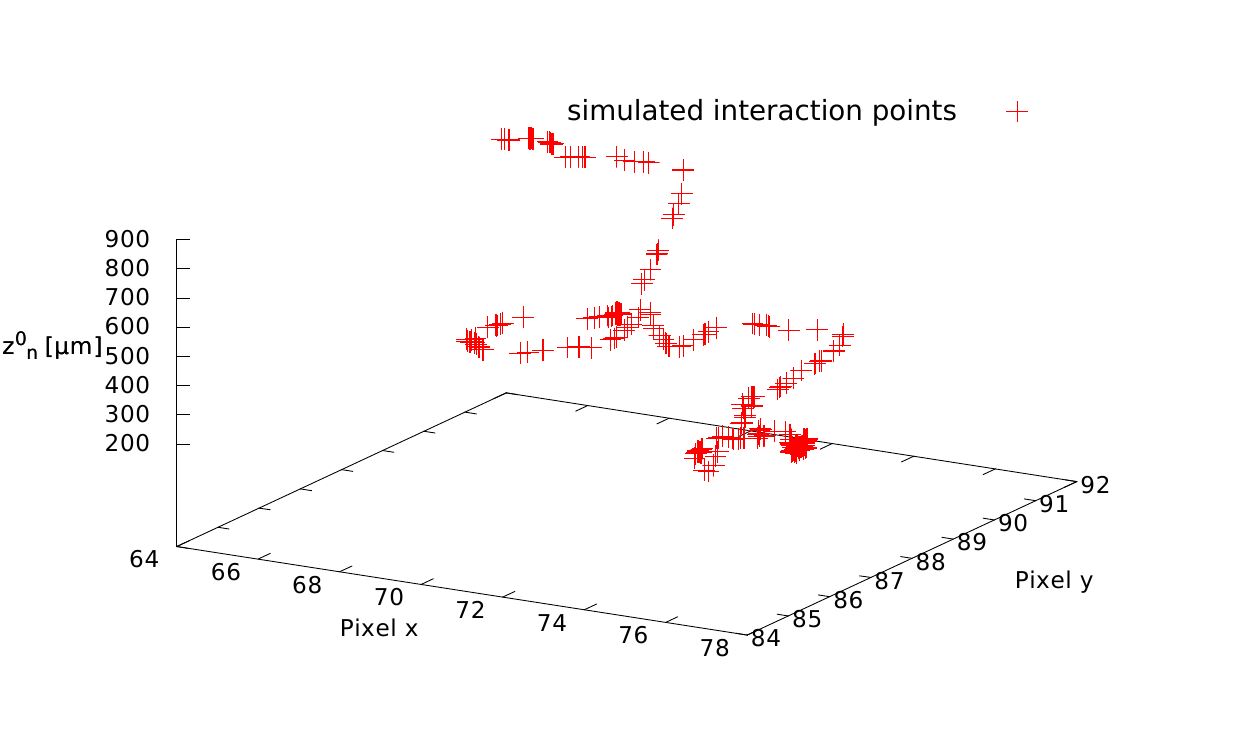}\label{sim_compl_track}}

\subfloat[Reconstructed Track]{\includegraphics[width=\columnwidth]{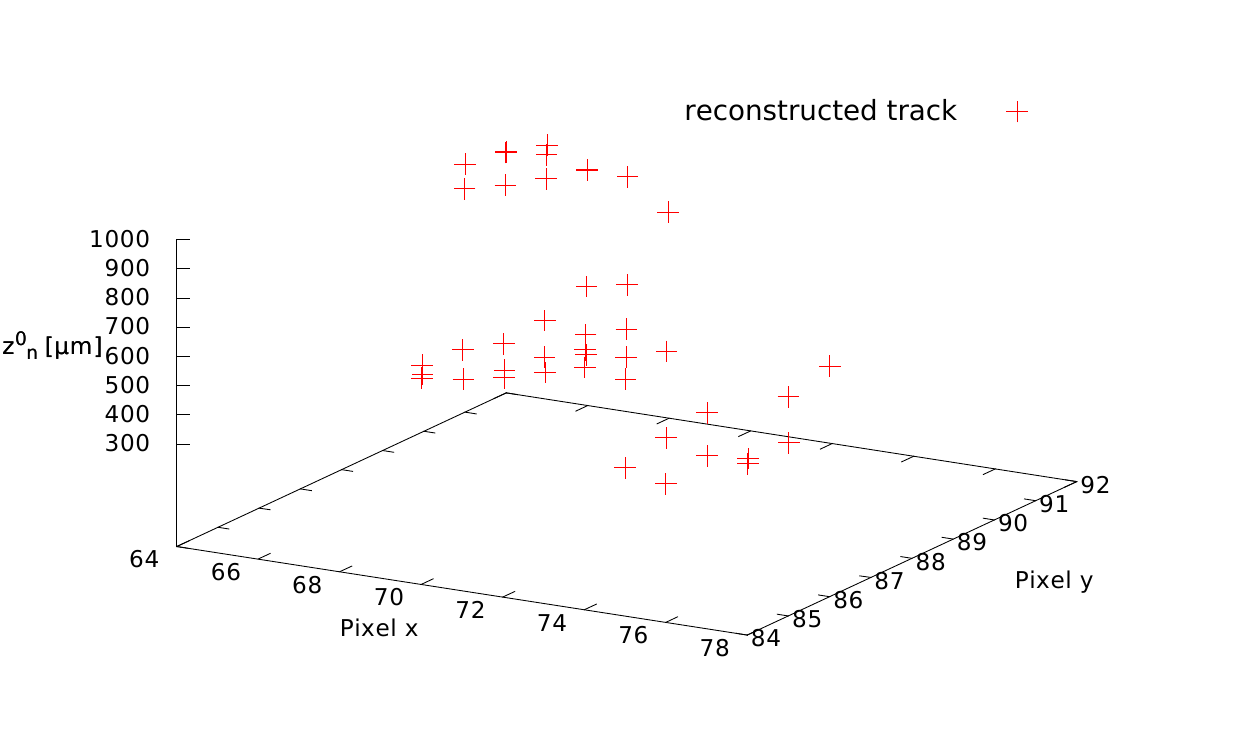}\label{sim_rec_track}}
\caption{Example of a simulated track (with all interaction points) and the reconstruction of this track obtained from pseudo-measured data.}
\end{figure}

Based on the physics and the method described in the previous subsection a detector has to be able to measure the following quantities for a 3D track reconstruction:

\begin{itemize}
	\item{The total charge induced at the cathode $Q_c$.} 
	\item{The charge released over every pixel $Q^{dep}_n$.}
	\item{The time-of-arrival of the charge signal in every pixel $\Delta t^{ToA}_n$.}
\end{itemize}

\begin{figure}[tb]
\subfloat[Timing Resolution]{\includegraphics[width=\columnwidth]{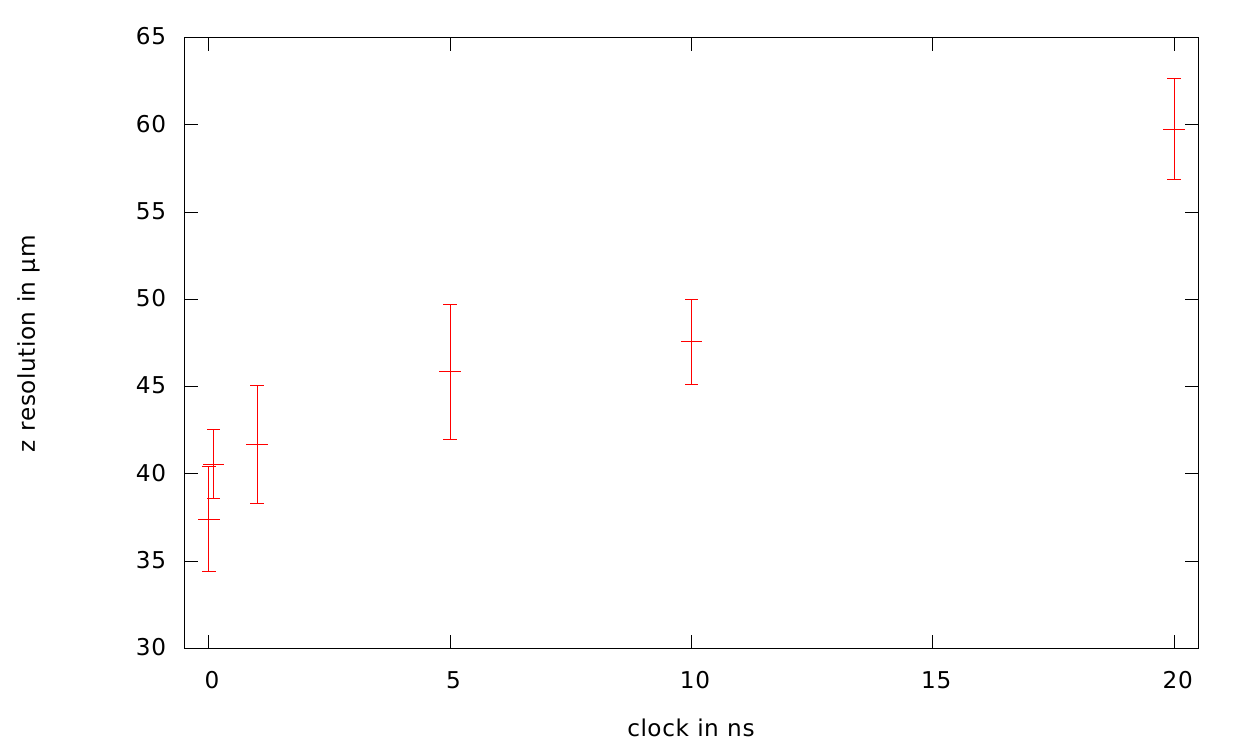}\label{z_res_timing}}

\subfloat[Energy Resolution]{\includegraphics[width=\columnwidth]{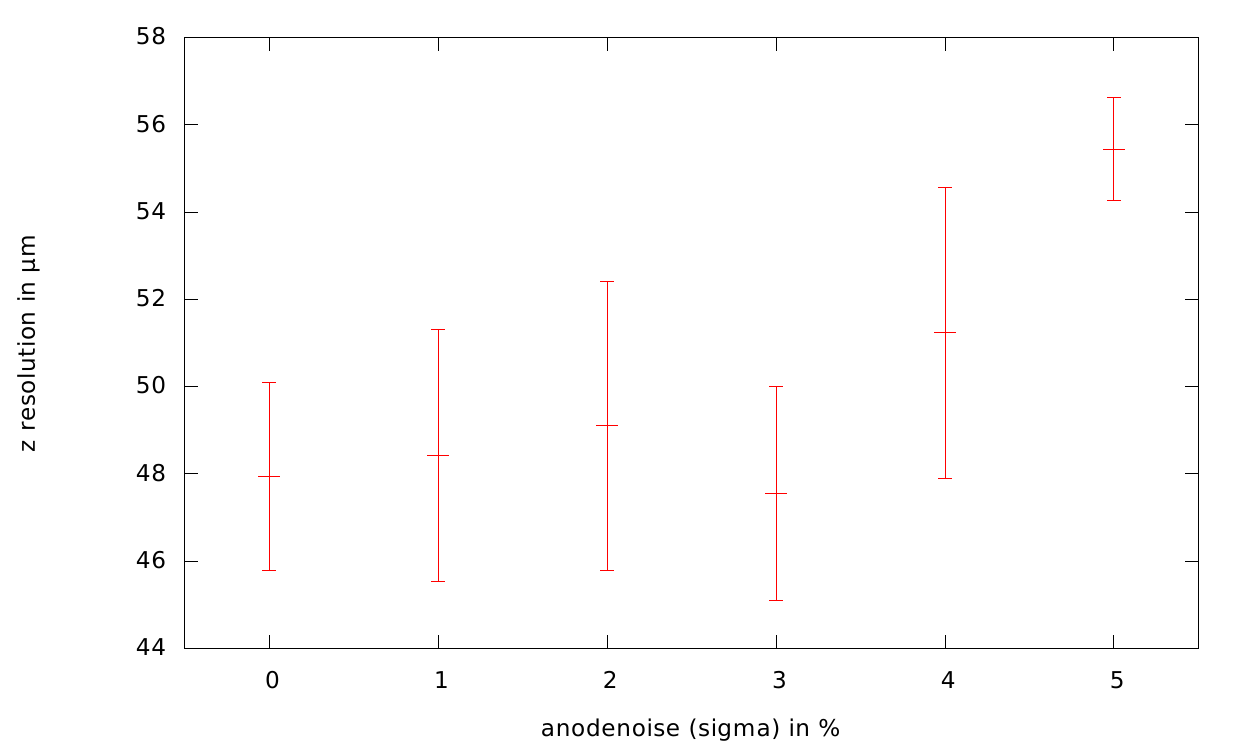}\label{z_res_energy}}
\caption{The interdependence between the z-position resolution and the timing resolution (a); and the energy resolution (b). The error bars are the given by the standard deviation if the simulation is repeated several times with the same configuration.}
\end{figure}

Based on this information, the calibration of the detector ($Q^{THL}_n$) and the material properties for the sensor layer ($\mu$, $E_z(U,z)$, $W_{pot}(z)$) the reconstruction algorithm opperates as following:

\begin{enumerate}
	\item{Compute numerically $z(t)$ for a given field configuration $E(U,z)$ in the sensor layer.}
	\item{Fix an arbitrary value for $\Delta t_s$ and vary $z_0$ for every pixel under the condition in Eq. \ref{for_z0condition}. In practise, it is reasonable to choose a value between 0 and $t_s - t_{max}$ for $\Delta t_s$ (for instance $\frac{t_s - t_{max}}{2}$), where $t_{max}$ is the drift time from the top of the sensor layer to the bottom.}
	\item{Determine the real average depth of interaction for the track from the $Q_c$/$Q_a$ ratio and the lookup-table.}
	\item{Vary $\Delta t_s$ under the condition given by Eq. \ref{for_zavgcondition} to reconstruct the correct positions $z^0_n$ for every pixel.}
\end{enumerate}

We tested our reconstruction method on data produced by the in-house developed Monte-Carlo simulation named ROSI \cite{rosi}, \cite{rosi2}. It is based on EGS4 and has a low energy extension with the interaction code LSCAT. For each event the simulation propagtes the corresponding particle through the sensor and calculates charge deposition in the sensor layer. For an electron of 2.8 MeV it produces about 5000 to 10000 interaction points along the track. Afterwards the drift of the secondary electrons and the signal generation in the pixel electrodes is simulated. The simulation takes into account the diffusion and recombination properties of CdTe as well as the repulsion of the charge under its electric field.\par

In Fig. \ref{sim_compl_track} the complete track (with all interaction points) of a 2.8 MeV electron is shown and in Fig. \ref{sim_rec_track} the reconstructed track. The reconstruction uses the quantities that would be obtained with a detector of 110 \textmu m pixel size, a 1 mm thick sensor layer with 44.8 V applied bias voltage, an energy threshold of 5 keV and a timing resolution of 10 ns in every pixel. \par
The position resolution in z-direction depends on the charge resolution (which is equivalent to the energy resolution) of the cathode and anode signals as well as on the timing resolution. Figure \ref{z_res_timing} shows the dependence of the z-resolution $FWHM_z$ on the timing resolution if the energy resolution ($\frac{\sigma}{E} = 2\%$) is fixed. Figure \ref{z_res_energy} illustrates how the $FWHM_z$ depends on the energy resolution under a fixed timing resolution (10 ns).

\section{Experimental Results}

For an experimental test of our method we performed two different experiments. For both experiments we used a Timepix detector with a 1mm thick Cadmium-Telluride sensor. The detector was bump-bonded and assembled by X-Ray Imaging Europe GmbH. The energy threshold of our particular detector was about 7 keV. We used the detector at a bias voltage of 44.8 V. The reason for this is to increase the drift time of the electrons to preferably high values. In contrast to the usual application of imaging where the drifttime should be as low as possible for fast couting, we need it to be rather long. The maximum clock frequency that can be used is 100 MHz which corresponds to pulses of 10 ns length. At the usual voltage (500 V) the drifttime is about 20 ns which is faster than the rising time after the preamplifier (about 80 ns). We need the electron drifttime to be significantly larger than that in order to measure drift time differences between individual pixels. According to our simulation the drifttime at 44.8 V is about 200 ns . We chose a bias of about 45 V but not lower since the charge collection efficiency saturates at about 40 V to 50 V. Hence, the chosen voltage is a reasonable tradeoff between a sufficient charge collection efficiency for a good signal in each pixel and a large drifttime for a "`time-of-flight"' position reconstruction. \par
Before the measurements we performed a global ToT calibration for the detector. The method of calibration is described in \cite{calibpub}. We used two calibration energies ($59.56 \textrm{ keV}$ from $^{241}\textrm{Am}$ and $80.99 \textrm{ keV}$ from $^{133}\textrm{Ba}$).

\subsection{Reconstruction of $\alpha$-particle Energy Depostion Distributions}

\begin{figure}[tb]

\subfloat[ToT]{\includegraphics[width=0.45\columnwidth]{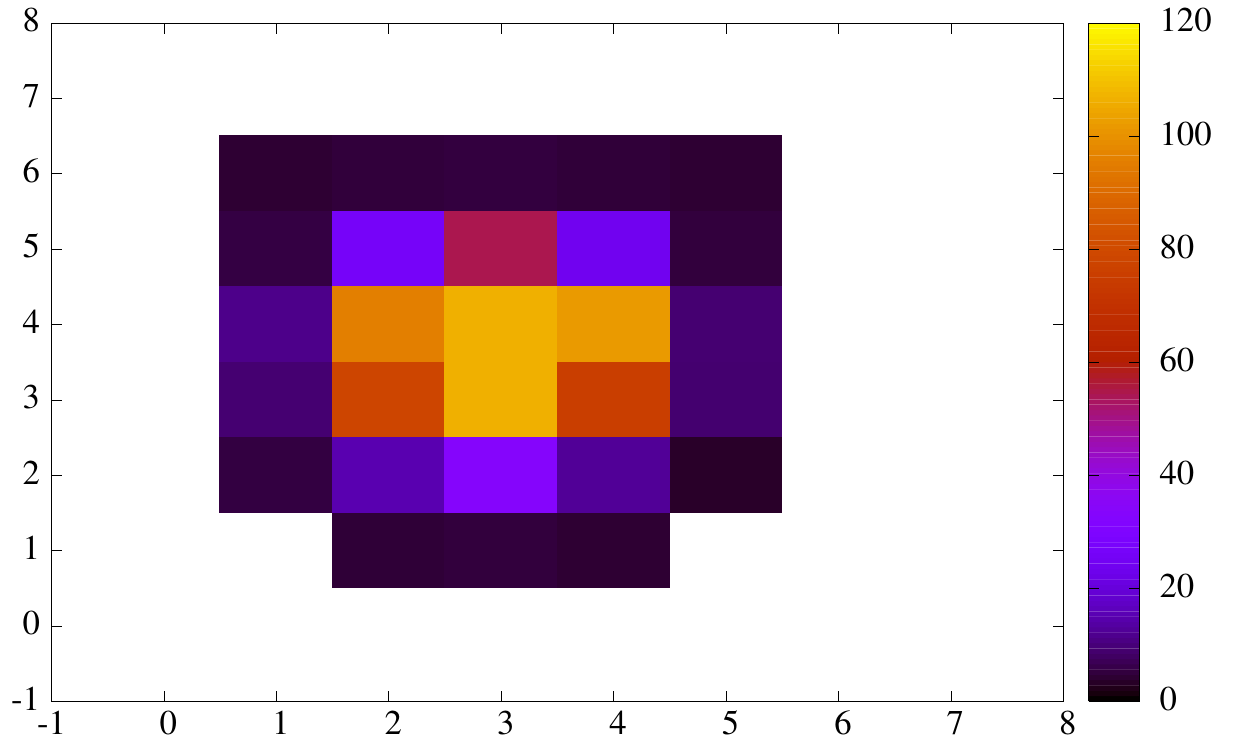}}

\subfloat[ToA]{\includegraphics[width=0.45\columnwidth]{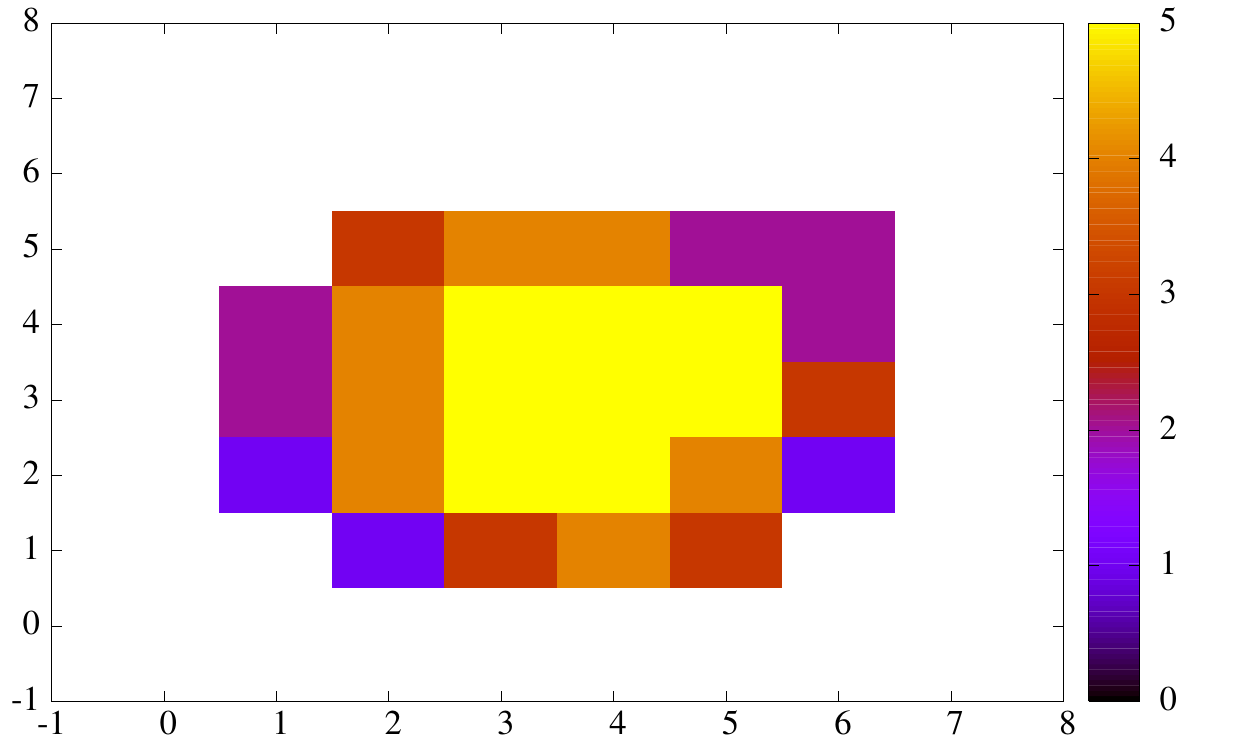}}
\subfloat[Energy from ToA]{\includegraphics[width=0.45\columnwidth]{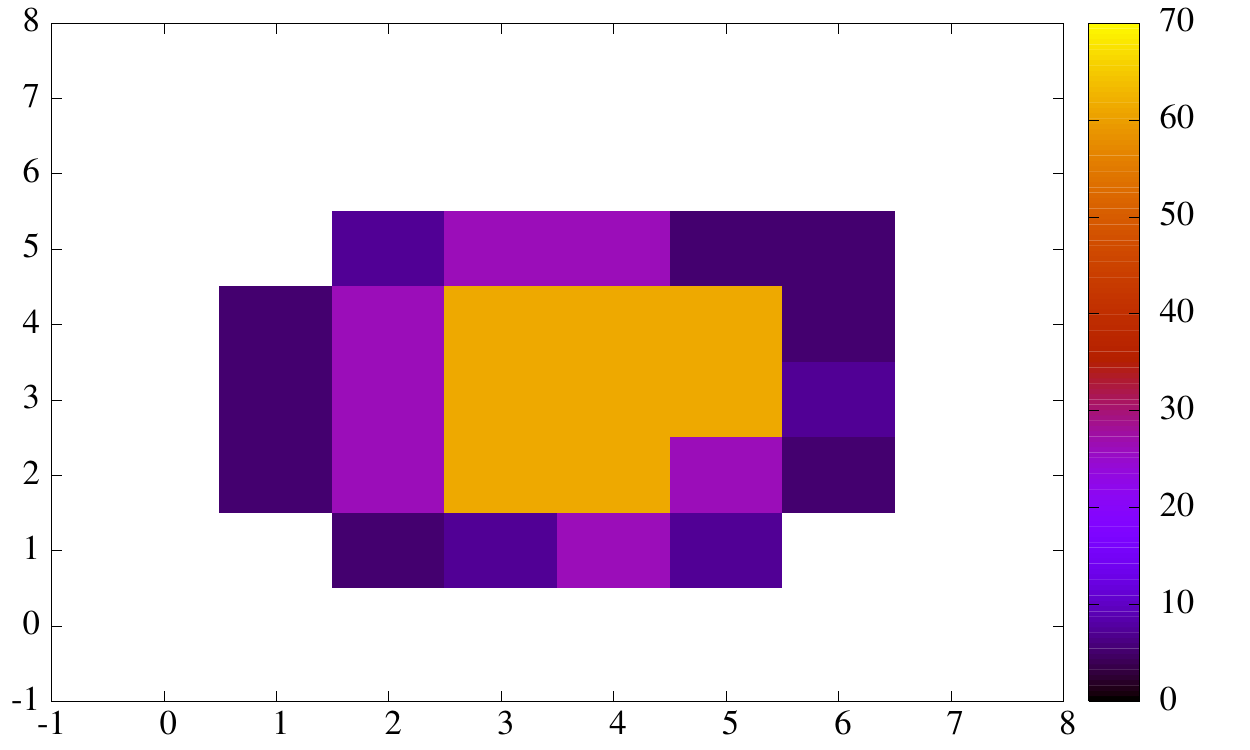}}
\caption{A typical $\alpha$-particle event measured in a) ToT mode; b) ToA mode. Image c) shows the event with energy reconstructed from ToA data. The color denotes the ToT, ToA and reconstructed energy, respectively.}
\label{alpha_events}
\end{figure}

Originally, in our method the z-position (depth of interaction) for every pixel triggered in an event is reconstructed from the simultaneous measurement of the total deposited energy, the energy deposited in every pixel and a timing information in every pixel. However, according to Eq. 5 the method can also be adopted to reconstruct the energy deposition in every pixel from the timing information, the total deposited energy and the knowledge of the z-positions. In this case the quantity $Q^{ind}_n$ (which is equivalent to the deposited energy) for every pixel is calculated numerically in 30 iterations with Eq. \ref{for_z0condition}. \par

With the Timepix it is impossible to obtain all of the three quantities simultaneously. Nonetheless, in order to show the functionality of the method, it was possible to overcome this disadvantage in the following way: We placed a $^{241}\textrm{Am}$ source with an activity of about $310 \textrm{ kBq}$ close to the sensor surface. The source delivers $\alpha$-particles that deposit a particular energy within the first 15 \textmu m of the sensor layer. Therefore, within good accuracy $z^0_n$ = 0 can be used for all pixels in the reconstruction which removes one parameter.\par

\begin{figure}[tb]
\includegraphics[width=\columnwidth]{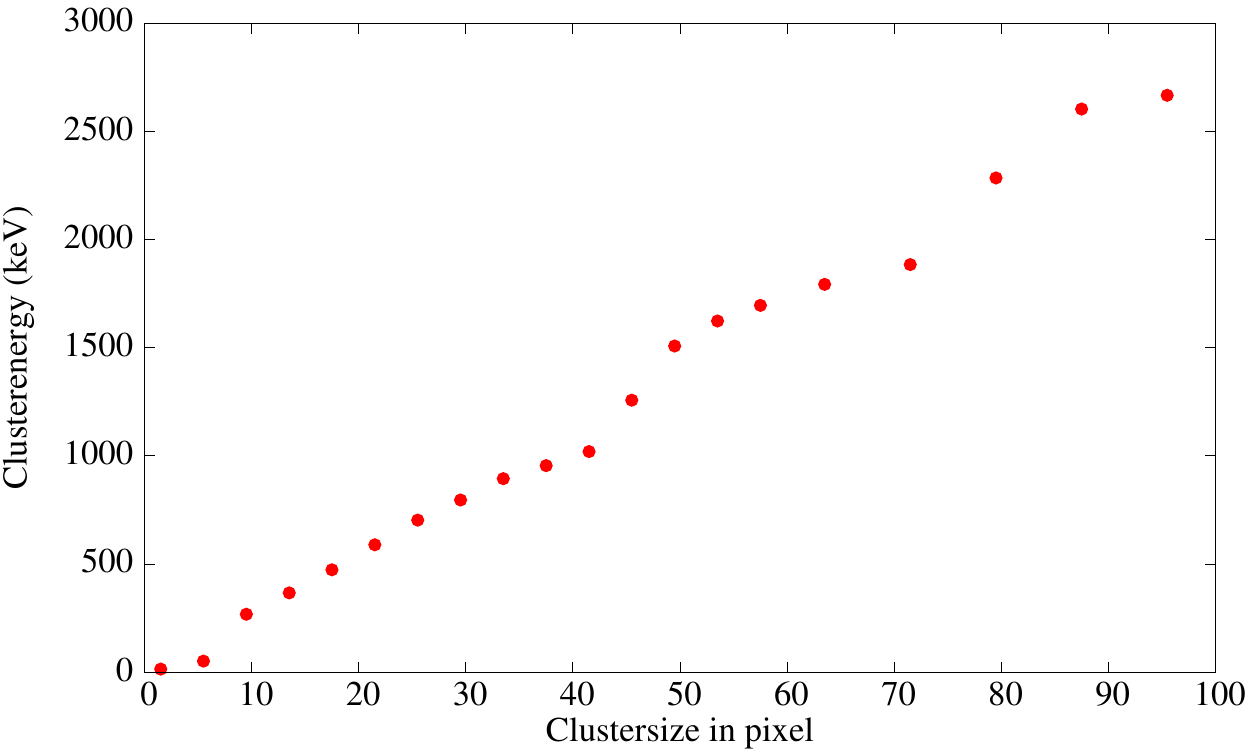}
\caption{The correlation between energy and cluster size from $\alpha$-particle events.}
\label{alpha_energy_size}
\end{figure}

First, we measured the energy deposition distribution per pixel in the ToT mode. A typical event cluster in ToT mode is shown in Fig. \ref{alpha_events} (a). The colour denotes the deposited energy in keV. By summing up all the energy values in the cluster we obtained the total deposited energy. The total deposited energy is linearly correlated with the cluster size (the number of triggered pixel in an event) as shown in Fig. \ref{alpha_energy_size}. As the cluster size is not affected by the mode of operation, we can use this correlation to determine the total deposited energy for $\alpha$-events measured in the ToA mode.\par

As next, we recorded a similar number of events in the ToA and the ToT mode. A typical event in ToT, ToA and energy reconstructed from ToA are shown in Fig.\ref{alpha_events} a),b) and c), respectively. Apparently, the dynamic range in the ToA data is lower since the total drift-time across the sensor layer is 200 ns and the clock that we used for this experiment was 48 MHz. Therefore, the maximum number of time slices is limited to 5. \par

\begin{figure}[tb]
\includegraphics[width=\columnwidth]{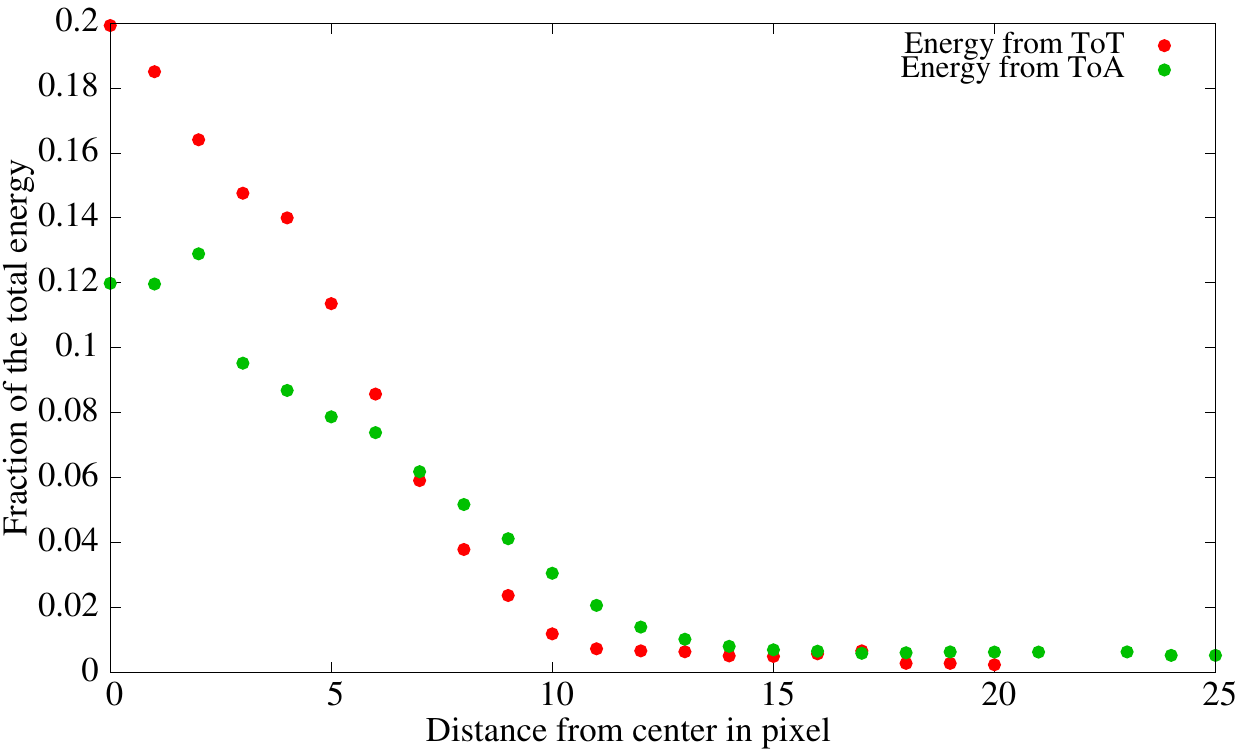}
\caption{The distribution of energy in an $\alpha$-event obtained from ToT data (red) and reconstructed from ToA data (green). The curves are averaged over many events.}
\label{alpha_energy_dist}
\end{figure}

A comparision between the energy distribution obtained in the ToT mode and reconstructed from the data in ToA mode is shown in Fig. \ref{alpha_energy_dist}. The fraction of energy deposited in a pixel is plotted against the distance from the center of the event. The discripancy has multiple reasons: In the central region of the cluster the amount of released charge is so high that the induced charge goes over the threshold very quickly and the drift time differences cannot be resolved with a resolution of 21 ns. Additionally, the charge is distributed among the pixels by charge sharing but we treat them as if the charge was initially released in the pixels. The result of this experiment suggests that the method has to be improved in the case of charge sharing but also highlights that the method works in principle. 

\subsection{Reconstruction of Electron Tracks}

\subsubsection{Experimental Design}

The second experiment was intended to demonstrate that 3D tracks can be reconstructed. Again, as the Timepix in its current version cannot measure all neccesary quantities at the same time, we designed our experiment in such a way that the parameters that cannot be measured by the detector directly, can be accessed and determined otherwise.\par

We used the Timepix in a so-called mixed matrix mode and took advantage of the fact that minimal ionizing particles deposit roughly the same energy in every pixel along its track. The Timepix allows to assign every pixel individually a mode of operation either ToA or ToT. For our experiment we chose the modes in a checkerboard arrangement - every pixel has an diagonal neighbour that is in the same mode and an off-diagonal neighbour that is in the different mode. It is a reasonable approximation to assume that minimal ionzing particles deposit nearly the same energy in every pixel. Hence, one can determine the energy deposited in a pixel that runs in the ToA mode by calculating the average value of deposited energy in the surrounding triggered ToT pixels. The second advantage of minimal ionizing particles is that they propagate through the complete sensor on an almost straight line. Thus, we know that the average z-position of the track is $\avg{z}^{meas}= \frac{d}{2}$ without measuring the total charge at the common electrode. \par 

Another way to determine the averaged deposited energy in every pixel is to take data in ToT mode and afterwards without changing the angle between the beam and the sensor layer to take data in ToA mode. From the ToT measurement we can determine the energy deposition in every pixel and afterwards the ToA data can be used for z-position reconstruction under the assumption that approximately the same energy was deposited in every pixel for each measurement in ToT and ToA mode under a fixed angle.\par

\begin{figure}[tb]

\subfloat[Mixed Mode]{\includegraphics[width=0.5\columnwidth]{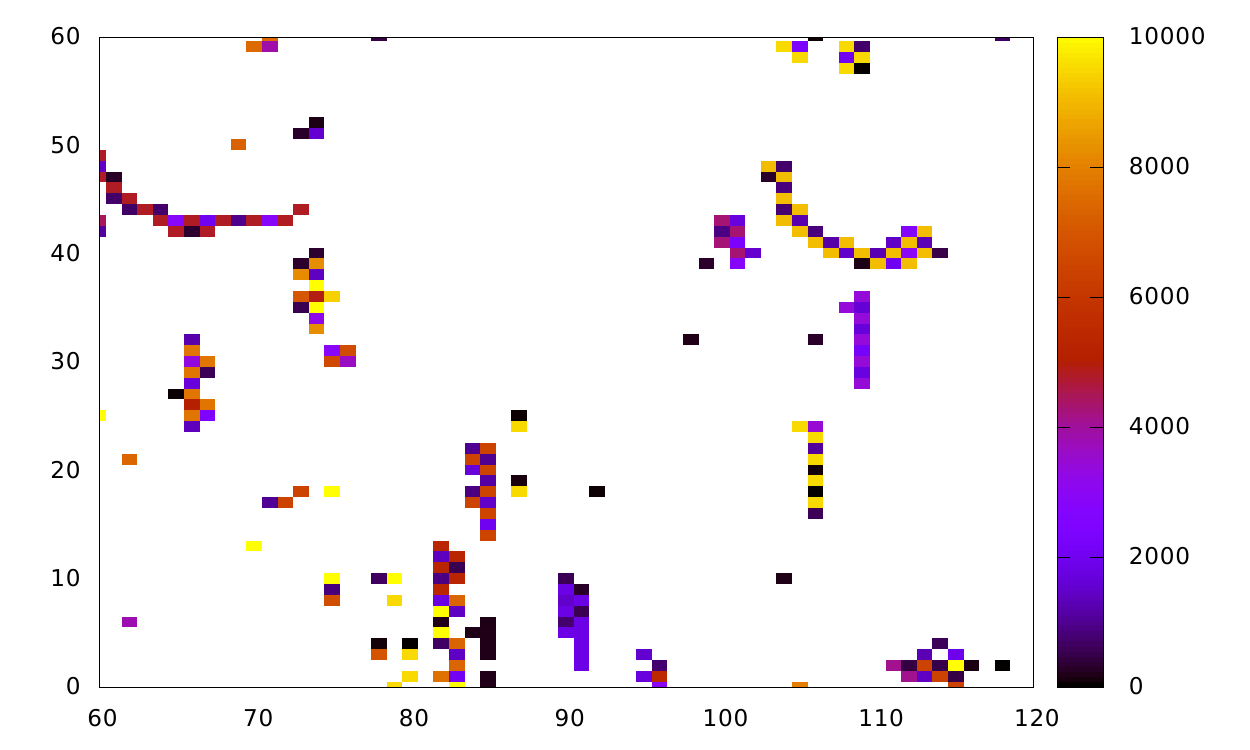}\label{el_events_mixed_beforeselec}}
\subfloat[Mixed Mode]{\includegraphics[width=0.5\columnwidth]{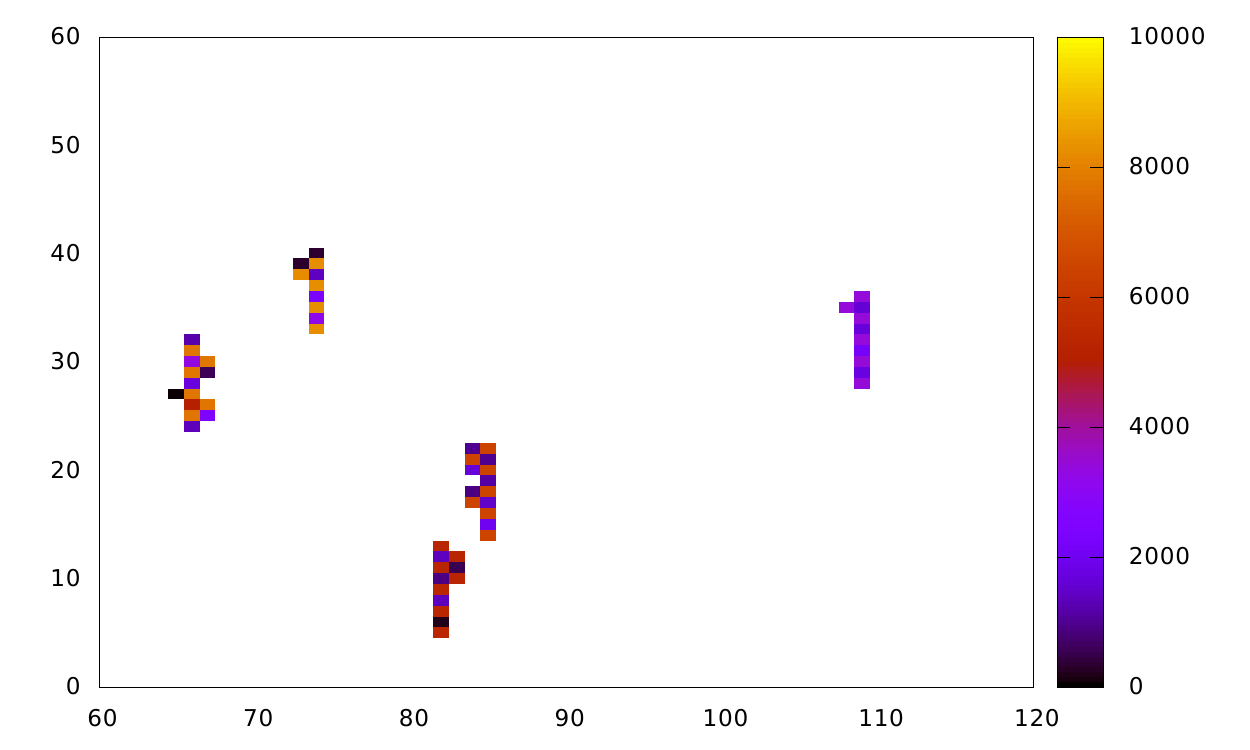}}

\subfloat[ToA Mode]{\includegraphics[width=0.5\columnwidth]{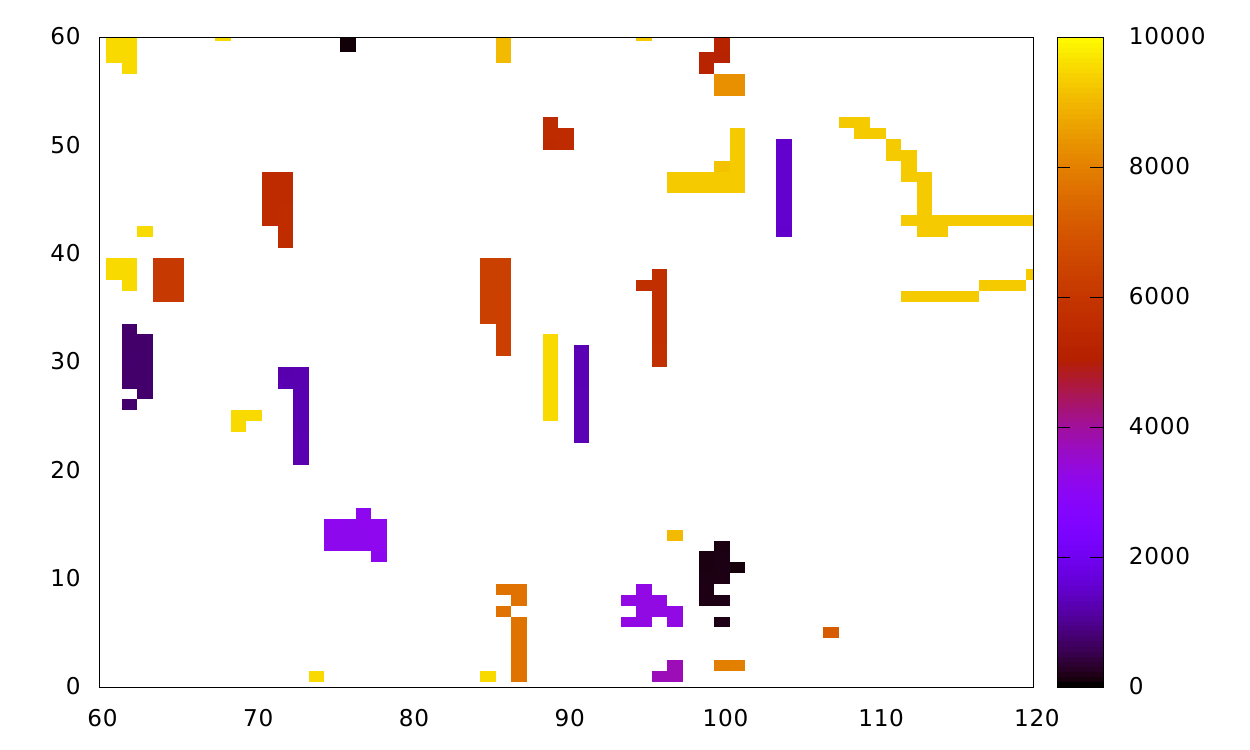}}
\subfloat[ToA Mode]{\includegraphics[width=0.5\columnwidth]{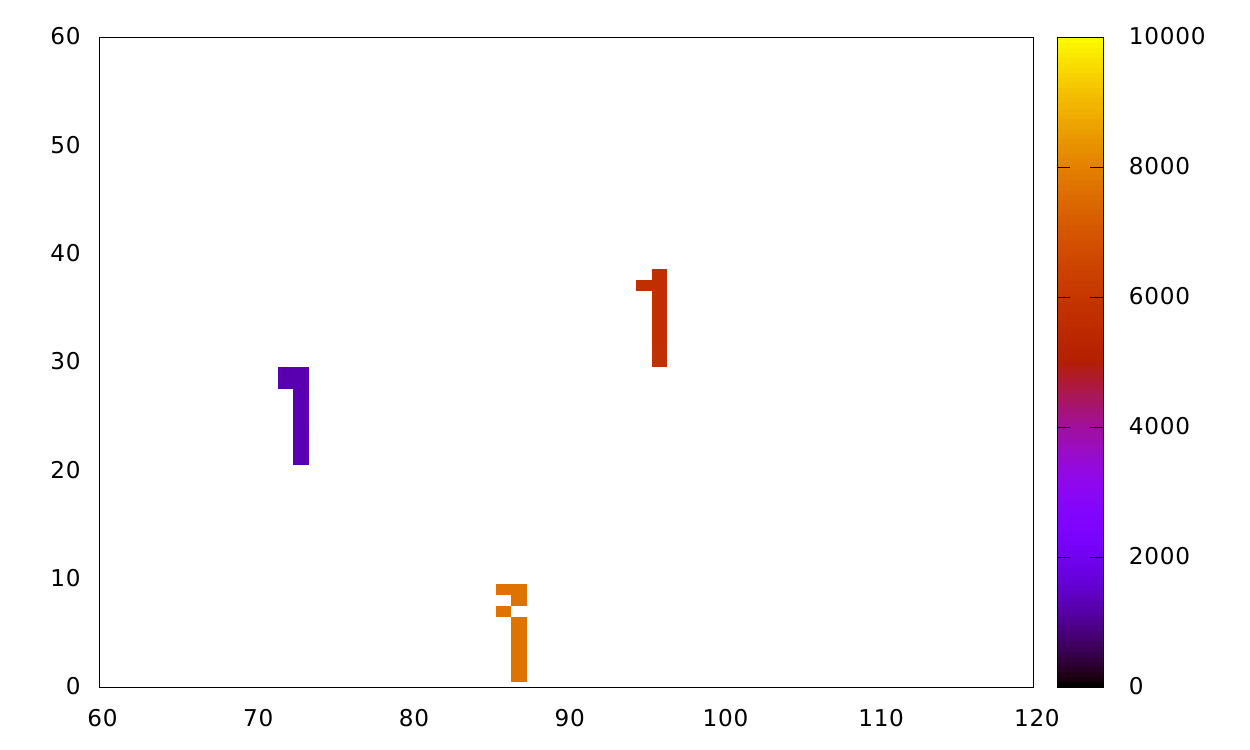}\label{el_events_toa_afterselec}}

\caption{Typical tracks for the evaluation of the z-resolution before (a,c) and after (b,d) track selection in the mixed-mode (a,b) and in the ToA mode (c,d). The colour bar denotes the measured ToA/ToT in every pixel.}
\end{figure}

\begin{figure}[tb]
\subfloat[Short Track]{\includegraphics[width=\columnwidth]{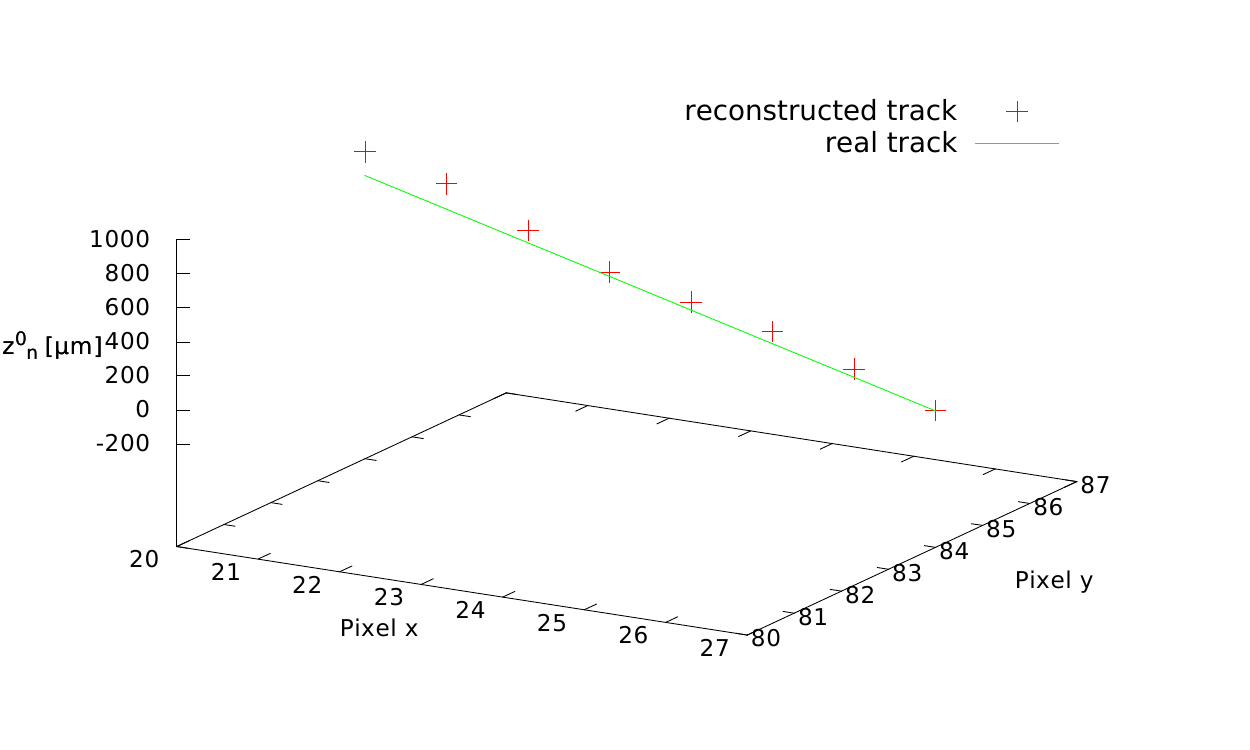}}

\subfloat[Long Track]{\includegraphics[width=\columnwidth]{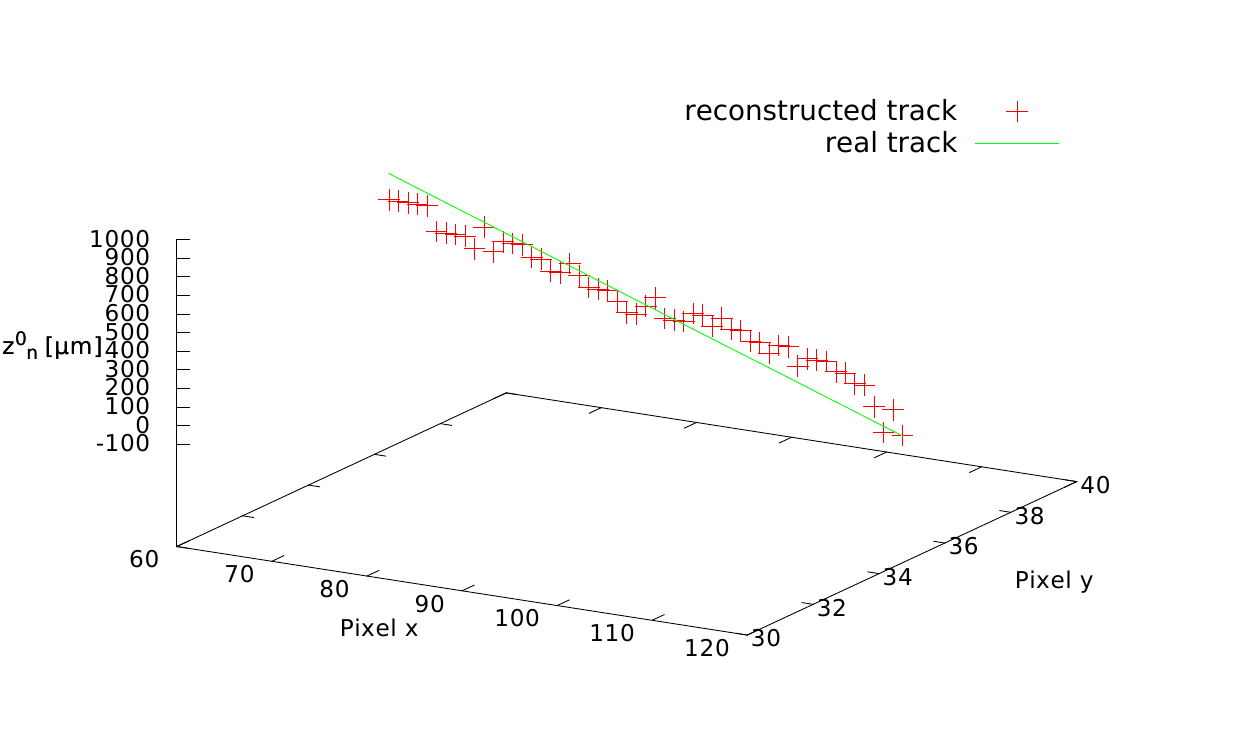}}
\caption{A reconstructed track of a typical short and long event after reconstruction in comparision to the real track.}
\label{reconstr_tracks}
\end{figure}

The data with minimal ionizing particles was aqcuired at the DESY T21 testbeam. We used an electron beam at 4.4 GeV. The sensor layer was positioned under two different angles to the beam direction. In one case the angle was about 42\textdegree$\ $and we obtained short tracks of 8 to 9 pixels in average and in the other case (10\textdegree) the tracks were 55 to 56 pixels long in average. The frametime was short (120 \textmu s) wherefor two tracks never overlap on one frame. We took data in the mixed, the ToT and the ToA mode.\par

\subsubsection{Data Analysis and Results}

For the evaluation of the resolution that we achieved in z-direction we performed a selection of tracks and sorted out tracks which are not meaningful due to the following criteria: First, we removed tracks where all pixel counted a ToA  of 11280. This is the maximum (hardware limited) value of the counter and if it is reached, we cannot extract any further information from the pixel (the true value could be any value bigger than 11280). Secondly, the electron beam was in x-direction while the detector was placed in the yz-plane and then tilted along the y-axis. Thus, as the tracks from minimal ionizing events are straight lines, meaningful events should trigger pixels mostly in one x-column. For events which do not produce straight lines due to scattering in the sensor our reconstruction will give false results since the assumption $\avg{z}^{meas} = \frac{d}{2}$ is not fulfilled. Hence, we used the following additional criterion for event selection: Only tracks which had twice as much triggered pixels in one x-column than in other columns combined are taken into account. Since hardly any long tracks fulfilled this criteria, we used only those for the further analysis. At last, we used only tracks with at least 10 trigger pixels to avoid background from low energy events. Figures \ref{el_events_mixed_beforeselec} to \ref{el_events_toa_afterselec} show a sample of tracks before and after event selection in the mixed-mode and in the ToA mode, respectively.\par

The experiment with $\alpha$-particles showed that our algorithm can give unsatisfying results for charge sharing pixels. Therefore, we removed pixels which are not in the main x-column as those are probably due to charge sharing. Also, we removed pixels with the highest y-coordinate values as these are most probably due to charge sharing from the pixels where the electron entered and left the sensor layer.\par
A typical track in comparision to the expected track for each case (long and short) is shown in Fig. \ref{reconstr_tracks} (a) and (b), respectively. At this point it is important to mention that the algorithm is not biased in any way to reconstruct a straight line. The fact that the result of reconstruction is a straight line indeed is a strong indicator that our method works properly. The reconstructed points are scattered close around the actual particle trajectory which indicates a meaningful result. \par
\begin{figure}[tb]
\subfloat[Mixed Mode]{\includegraphics[width=\columnwidth]{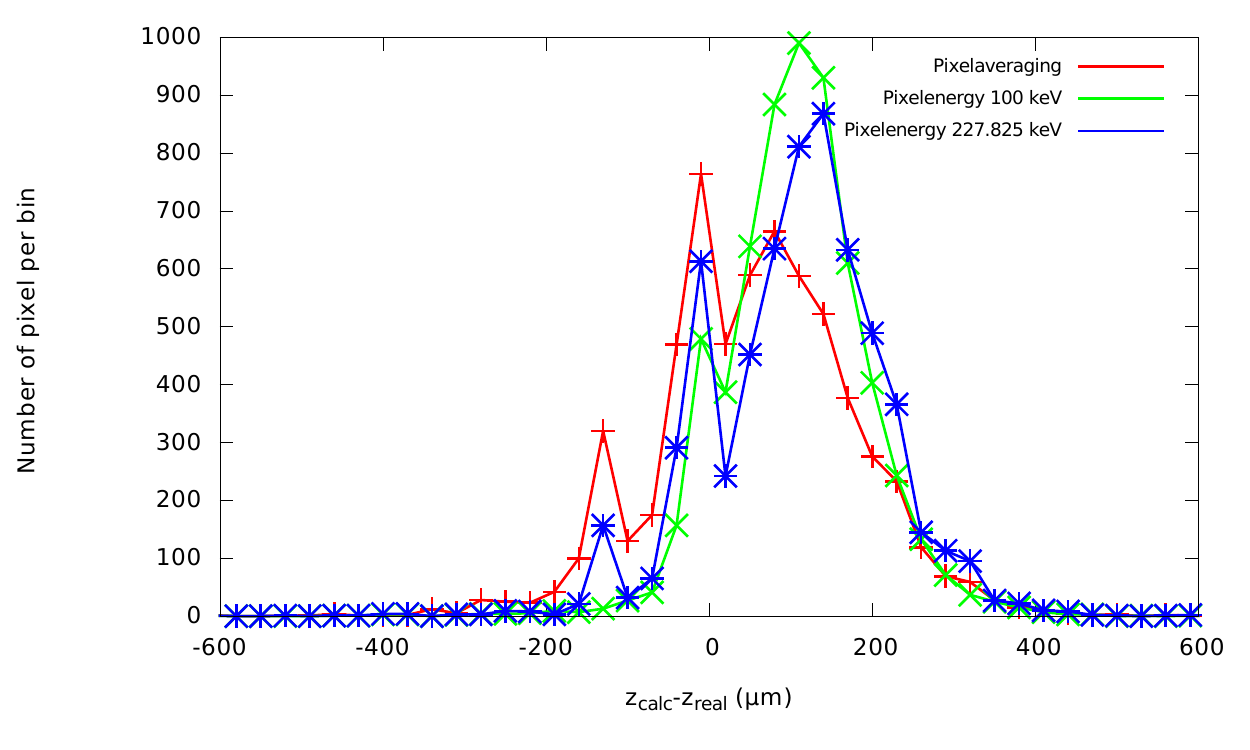}\label{z_res_mixed}}

\subfloat[ToA Mode]{\includegraphics[width=\columnwidth]{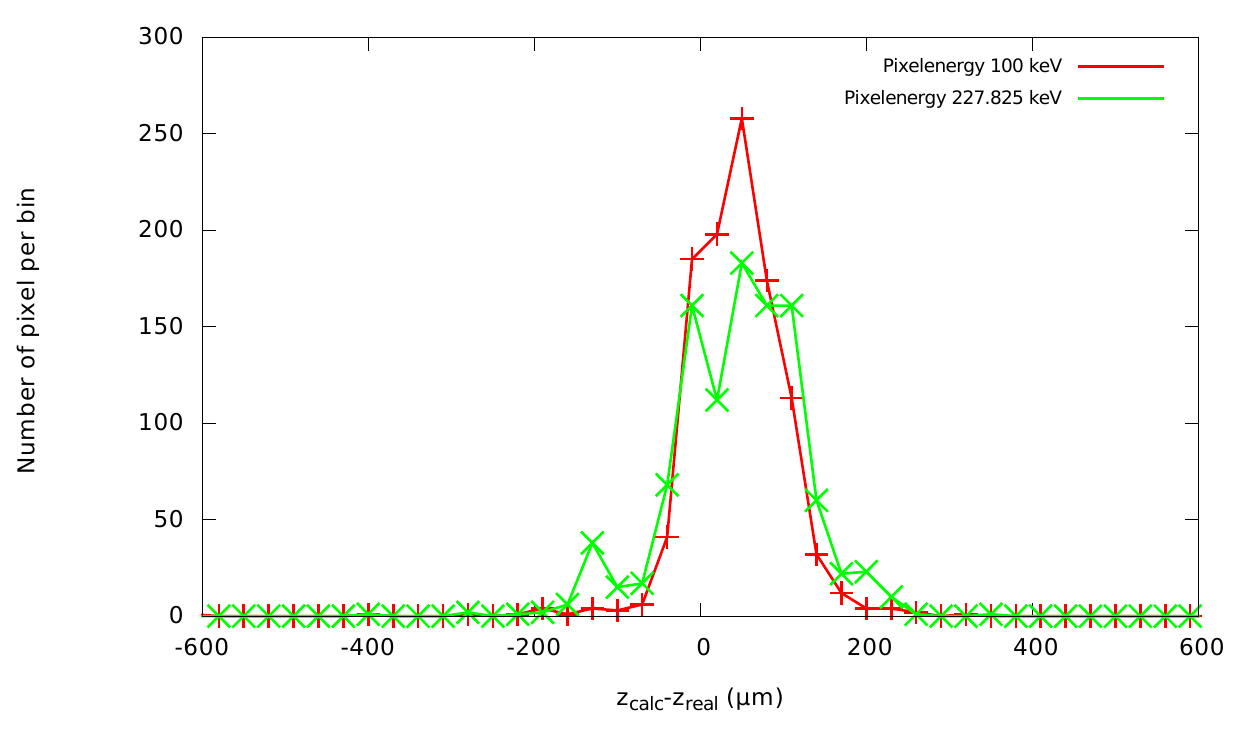}\label{z_res_toa}}
\caption{The deviations of the reconstructed z-position from the actual track for events measured in mixed mode (a) and ToA mode (b).}
\end{figure}

We calculated the z-resolution as the quadratic sum of deviations of the reconstructed z-position in every pixel from a straight line between the pixels with the highest and lowest y-coordinate values normalized to the number of pixels. This line is what comes closest to the real particle trajectory. The results are summerized in table \ref{tab_resolution}. Figure \ref{z_res_mixed} and \ref{z_res_toa} show the deviations of the reconstructed z-position from the actual track.\par

\begin{table}[tb]
\centering
\begin{tabular}{r|cc}
& Mixed Mode & ToA Mode \\
\hline
Neighbours Averaging & 141 \textmu m & 99 \textmu m \\
All Pixel at 227 keV & 160 \textmu m & 149 \textmu m \\
All Pixel at 100 keV & 149 \textmu m  & 84 \textmu m \\
\hline
\end{tabular}
\caption{Results on the z-resolution obtained in different modes of operation.}
\label{tab_resolution}
\end{table}

The experiments were limited to straight tracks from minimal ionizing particles since the available detectors cannot provide all the information required for reconstruction. Nonetheless the next generation of Timepix detectors (Timepix3) \cite{tp3} will provide a simultaneous readout of timing and energy deposition information per pixel. Also the readout of a backside signal could be included as for a coplanar-grid-type of detector. With a proposed timing resolution of 1.6 ns, a voxel detector with about 50 \textmu m voxel size can be expected. A Timepix3 detector with a CdTe sensor could therefore provide an interesting opportunity for new imaging techniques with voxel detectors.\par

In the mixed-mode we achieved an energy resolution of 141 \textmu m. By using the data from ToA only and a fixed energy of $227 \textrm{ keV}$ we achieve a z-resolution of 99 \textmu m. This value for the average energy was determined from the ToT measurement as the average deposited energy per pixel. However, as this energy is a constant for all pixels in the reconstruction, it can also be set to a different fixed value. It turns out that the best resolution of 84 \textmu m could be achieved by using the data from ToA mode and a fixed energy of $100 \textrm{ keV}$ for every pixel.

\section{Conclusions and Outlook}

In this paper we presented and evaluated a method to reconstruct 3D particle trajectories through a semiconductor sensor layer. Our simulations suggests that a z-resolution of about 40 to 50 \textmu m can be achieved under realistic chip performance assumption. \par
As a major result, we successfully demonstrated that the method works experimentally and a particle track can be reconstructed from actual data. The achieved z-resolution was 84 \textmu m in the best case.\par

\section*{Acknowledgement}

We would like to thank Dominik Dannheim and Ralf Diener for their help and giving us the opportunity to acquire data at the DESY test beam facility. Also, we would like to thank the Medipix Collaboration for supporting this work.

\bibliographystyle{spphys}       

\bibliography{references}

\providecommand{\noopsort}[1]{}\providecommand{\singleletter}[1]{#1}%
\begin{thebibliography}{10}
\providecommand{\url}[1]{{#1}}
\providecommand{\urlprefix}{URL }
\expandafter\ifx\csname urlstyle\endcsname\relax
  \providecommand{\doi}[1]{DOI \discretionary{}{}{}#1}\else
  \providecommand{\doi}{DOI \discretionary{}{}{}\begingroup
  \urlstyle{rm}\Url}\fi

\bibitem{majorana}
E.~Majorana, Nuevo Cimento \textbf{14} (1937)

\bibitem{dbreview}
P.~Vogl, S.~Elliott, Ann. Rev. Nucl. Part. Sci. \textbf{52}, 115 (2002)

\bibitem{kailbnl}
B.~Plimley, K.~Vetter, et~al., Nucl. Instr. Meth. A \textbf{654}, 244 (2011)

\bibitem{msu}
J.~Kim, Z.~He, et~al., Nucl. Instr. Meth. A \textbf{683}, 53 (2012)

\bibitem{physrevbenni}
B.~Bergmann, T.~Michel, et~al., Phys. Rev. C \textbf{89} (2014)

\bibitem{mpx}
X.~Llopart, M.~Campbell, et~al., Nucl. Instr. Meth. A \textbf{581}, 485 (2007)

\bibitem{fitpix}
M.~Platkevic, J.~Jakubek, Z.~Vykydal, et~al., Nucl. Instr. Meth. A
  \textbf{591}, 254

\bibitem{calibpub}
M.~Filipenko, T.~Gleixner, et~al., EPJ C  (2013)

\bibitem{hindawi}
T.~Michel, T.~Gleixner, et~al., AHEP \textbf{2013} (2013)

\bibitem{qindcalc}
W.~Shockley, J. Appl. Phys. \textbf{9}, 635 (1938)

\bibitem{weightpot}
A.~Castoldi, E.~Gatti, P.~Rehak, et~al., IEEE Trans. Nucl. Sci. \textbf{43}(3),
  256 (1996)

\bibitem{msuzpos}
W.~Li, Z.~He, et~al., IEEE Trans. Nucl. Sci. \textbf{47}(3), 890 (2000)

\bibitem{rosi}
J.~Giersch, J.~Anton, et~al., Nucl. Intsr. Meth A \textbf{509}, 151 (2003)

\bibitem{rosi2}
J.~Durst, J.~Giersch, Nucl. Instr. Meth. A \textbf{591}, 300 (2007)

\bibitem{tp3}
M.~van Beuzekom, F.~Zappon, M.~Campbell, X.~Llopart, et~al., in
  \emph{Proceedings of Science} (SISSA, 2011), pp. 1--8

\end{thebibliography}

\end{document}